\RequirePackage{fix-cm}
\documentclass[smallextended]{svjour3}       % onecolumn (second format)
\smartqed  % flush right qed marks, e.g. at end of proof
\usepackage{graphicx}
\usepackage{cite}
\usepackage{amsmath,amsfonts,amssymb}
\usepackage{subfigure}
\usepackage{float}
\usepackage{color}
\usepackage{multirow}
\usepackage{mathtools}
 \journalname{Advances in Computational Mathematics}
\begin{document}

\title{A stabilized proper orthogonal decomposition reduced-order model for large scale quasigeostrophic ocean circulation%\thanks{Grants or other notes
}

\titlerunning{Reduced-order modeling of barotropic vorticity equation}        % if too long for running head

\author{Omer San         \and
        Traian Iliescu %etc.
}


\institute{O. San \at
              Interdisciplinary Center for Applied Mathematics \\
              Virginia Tech, Blacksburg, VA 24061, USA \\
              Tel.: +1 (540) 231 5054\\
              Fax: +1 (540) 231 7079\\
              \email{omersan@vt.edu}           %  \\
           \and
           T. Iliescu \at
              Department of Mathematics \\
              Virginia Tech, Blacksburg, VA 24061, USA
}

\date{Received: \today }

\maketitle

\begin{abstract}
In this paper, a stabilized proper orthogonal decomposition (POD) reduced-order model (ROM) is presented for the barotropic vorticity equation. We apply the POD-ROM model to mid-latitude simplified oceanic basins, which are standard prototypes of more realistic large-scale ocean dynamics. A mode dependent eddy viscosity closure scheme is used to model the effects of the discarded POD modes. A sensitivity analysis with respect to the free eddy viscosity stabilization parameter is performed for various POD-ROMs with different numbers of POD modes. The POD-ROM results are validated against the Munk layer resolving direct numerical simulations using a fully conservative fourth-order Arakawa scheme. A comparison with the standard Galerkin POD-ROM without any stabilization is also included in our investigation. Significant improvements in the accuracy over the standard Galerkin model are shown for a four-gyre ocean circulation problem. This first step in the numerical assessment of the POD-ROM shows that it could represent a computationally efficient tool for large scale oceanic simulations over long time intervals.
\end{abstract}

\section{Introduction}
\vspace{-2pt}
\label{sec:intro}
Proper orthogonal decomposition (POD) is one of the most successful successful reduced-order modeling techniques of complex systems. POD has been used to generate reduced-order models (ROMs) for the optimal control and analysis of many forced-dissipative nonlinear systems in science and engineering applications \cite{ito1998reduced,iollo2000stability,noack2003hierarchy,rowley2006dynamics,bui2008model,hay2009local}. POD extracts the most energetic modes, which are expected to contain the dominant characteristics of these systems. The POD and its variants are also known as Karhunen-Lo\`{e}ve expansions in signal processing and feature selection \cite{fukunaga1970application}, principal component analysis in statistics  \cite{hotelling1933analysis,wold1987principal}, and empirical orthogonal functions in atmospheric science \cite{north1984empirical}. The development of accurate and reliable low dimensional models is crucial in many complex systems, such as data assimilation in weather and climate modeling \cite{daescu2007efficiency,cao2007reduced,daescu2008dual}.

Reduced-order modeling of such problems, usually governed by a system of coupled nonlinear partial differential equations, typically consists of a basis selection strategy to build representative modes and then a projection step to build the low-dimensional model. The globally supported POD modes are often constructed empirically from a database obtained from a high fidelity numerical simulation of the governing equations and are problem dependent. These bases are then used to reduce the partial differential equations to a truncated system of amplitude equations using Galerkin projection \cite{holmes1998turbulence}. It is possible to obtain good approximations with a few POD modes in which fine scale details are embedded. The resulting systems are low dimensional but dense and provide an efficient framework for real time analysis and control applications.

%These low-dimensional models are effectively obtained by considering the first few most energetic POD modes with Galerkin projection.
Although the POD-ROM Galerkin method provides an efficient way to generate the reduced-order system (especially for fairly smooth systems in which the energetics can be characterized by the first few modes), its applicability to complex systems is limited mainly due to errors associated with the truncation of POD modes. To model the effects of the discarded modes, several closure modeling strategies have been proposed \cite{couplet2003intermodal,kalb2007intrinsic,bergmann2009enablers,kalashnikova2010stability,wang2011two,carlberg2011efficient,wang2012proper,amsallem2012stabilization,balajewicz2013low,lassila2013model}.

%Several successful stabilization methods for POD applications have been suggested in order to model the effects of discarded POD modes \cite{kalb2007intrinsic,bergmann2009enablers,borggaard2011artificial,wang2011two,wang2012proper}. In that sense, conjectures in stabilization models for POD systems share in common with closure models of large eddy simulations of turbulent flows \cite{couplet2003intermodal}.

%The main motivation in this paper is to build a POD-ROM framework for the quasigeostrophic equations using a mode dependent eddy viscosity closure approach. The performance of this model is then analyzed for wind-driven forced-dissipative large scale ocean circulation problems. Solving two numerical examples of the four-gyre ocean circulation problem, we show that the stabilized POD-ROM model captures flow dynamics correctly whereas the standard POD-ROM with Galerkin method not.

The barotropic vorticity equation (BVE), also known as the single-layer quasigeostrophic model, is one of the most used mathematical models for forced-dissipative large scale ocean circulation problem.
Studies of wind-driven circulation using an idealized double-gyre wind forcing have played an important role in understanding various aspects of ocean dynamics, including the role of mesoscale eddies and their effect on the mean circulation.
The POD, along with other optimal bases choices, has been used to derive computationally efficient ROMs of the BVE (see, e.g., \cite{selten1995efficient,galan2008error}).
Both deterministic and stochastic closure schemes for the resulting POD-ROMs have been used (see \cite{crommelin2004strategies} for a survey).
The main goal of this paper is to investigate a mode dependent eddy viscosity closure model for POD-ROMs of the BVE.
Wind-driven forced-dissipative large scale ocean circulation problems for two different sets of physical parameters are used to test the closure model.
A standard Galerkin POD-ROM is also used for comparison purposes.
We note that, although sharing some features with the setting used in \cite{selten1995efficient}, the numerical investigation in this paper displays several significant differences.
The most important differences between the two settings are (i) the fundamentally different dissipation mechanism used in the BVE; (ii) the different physical parameters that yield completely different flow patterns; (iii) the different approaches used to generate the POD modes; (iv)  the different treatment of the differential operators in the POD-ROM closure models; and (v) the different numerical methods employed in the two investigations.

%(ii) the double-gyre wind forcing mechanism used in large scale oceanic circulation,

The organization of this paper is as follows. The BVE for large scale quasigeostrophic ocean model is summarized in Section \ref{sec:model}. The POD-ROM low-dimensional representation of the governing equations is presented in Section \ref{sec:rom}. The numerical schemes for the mathematical models are briefly described in  Section \ref{sec:num}. The results of the POD-ROM computations are compared with the Munk layer resolving direct numerical simulation (DNS) computations in Section \ref{sec:result} for solving the four-gyre wind-driven ocean circulation problems. Finally,  Section \ref{sec:conc} consists of summary and our concluding remarks.

%\begin{table}
%\caption{<Table caption>}
%\centering
%\tabsize
%\begin{tabular}{<table alignment>}
%\toprule
%<column headings>\\
%\midrule
%<table entries
%(separated by & as usual)>\\
%<table entries>\\
%.
%.
%.\\
%\bottomrule
%\end{tabular}
%\end{table}

\vspace{-6pt}

\section{Barotropic vorticity equation}
\vspace{-2pt}
\label{sec:model}
%In this section, we present the barotropic vorticity equation (BVE), one of the most used mathematical models for forced-dissipative large scale ocean circulation problem. It is also known as single-layer quasigeostrophic model.
Following \cite{san2011approximate}, we briefly describe the BVE. More details on the physical mechanism and various formulations can be found in \cite{gill1982atmosphere,pedlosky1982geophysical,vallis2006atmospheric,mcwilliams2006fundamentals,cushman2011introduction}.
%Studies of wind-driven circulation using an idealized double-gyre wind forcing have played an important role in understanding various aspects of ocean dynamics, including the role of mesoscale eddies and their effect on mean circulation.
The BVE for one-layer quasigeostrophic forced-dissipative ocean model can be written as
\begin{equation}\label{eq:bve}
\frac{\partial \omega}{\partial t} + J(\omega,\psi) -\beta\frac{\partial \psi}{\partial x} = D + F,
\end{equation}
where $D$ and $F$ represent the dissipation and forcing terms, respectively. In Eq.~(\ref{eq:bve}), $\omega$ is the kinematic vorticity, the curl of the velocity field, defined as
\begin{equation}\label{eq:vor}
\omega = \frac{\partial v}{\partial x} - \frac{\partial u}{\partial y},
\end{equation}
and $\psi$ is symbolizes the velocity stream function. The flow velocity components can be found from the stream function according to the following definitions:
\begin{equation}\label{eq:vel}
u = \frac{\partial \psi}{\partial y}, \quad v =- \frac{\partial \psi}{\partial x}.
\end{equation}
Thus, the kinematic relationship between the vorticity and stream function yields the following elliptic subproblem:
\begin{equation}\label{eq:poi}
\nabla^2 \psi = -\omega,
\end{equation}
where $\nabla^2$ is the two-dimensional Laplacian operator.
The BVE given by Eq.~(\ref{eq:bve}) uses the beta-plane approximation, which is valid for most of the oceanic basins. To account for the Earth's rotational effects, in the beta-plane approximation the Coriolis parameter is approximated by $f=f_0+\beta y$, where $f_0$ is the constant mean Coriolis parameter and $\beta$ is the gradient of the Coriolis parameter at the basin center (i.e., $y=0$). The nonlinear convection term in Eq.~(\ref{eq:bve}), called the nonlinear Jacobian, is defined as
\begin{equation}\label{eq:jac}
J(\omega,\psi) =  \frac{\partial \psi}{\partial y}\frac{\partial \omega}{\partial x} - \frac{\partial \psi}{\partial x}\frac{\partial \omega}{\partial y}.
\end{equation}
The viscous dissipation mechanism has the standard form
\begin{equation}\label{eq:vel}
D = \nu \nabla^2 \omega,
\end{equation}
where $\nu$ is the eddy viscosity coefficient. The double-gyre wind forcing function in the model is given by
\begin{equation}\label{eq:forc}
F = \frac{\tau_0}{\rho H} \frac{\pi}{L} \sin\Big(\pi \frac{y}{L}\Big),
\end{equation}
where $\tau_0$ is the maximum amplitude of the double-gyre wind stress, $\rho$ is the mean fluid density, and $H$ is the mean depth of the ocean basin. In order to obtain a dimensionless of the BVE,  we use the following definitions:
\begin{equation}\label{eq:dimless}
\tilde{x} = \frac{x}{L}, \quad \tilde{y} = \frac{y}{L}, \quad \tilde{t} = \frac{t}{L/V}, \quad \tilde{\omega} = \frac{\omega}{V/L}, \quad \tilde{\psi} = \frac{\psi}{V L},
\end{equation}
where the tilde denotes the corresponding nondimensional variables. In the nondimensionalization, $L$ represents the characteristic horizontal length scale (in our study $L$ is the basin dimension in the $x$ direction), and $V$ is the characteristic velocity scale. The Sverdrup velocity scale used for nondimensionalization can be written in the following form
\begin{equation}\label{eq:sverdrup}
V = \frac{\tau_0}{\rho H}\frac{\pi}{\beta L}.
\end{equation}
Finally, the governing equations for two-dimensional incompressible barotropic quasigeostrophic flows can be written in dimensionless form in beta-plane as the dimensionless BVE
\begin{equation}\label{eq:nbve}
\frac{\partial \omega}{\partial t} + J(\omega,\psi) -\frac{1}{Ro}\frac{\partial \psi}{\partial x} = \frac{1}{Re}\nabla^2 \omega + \frac{1}{Ro}\sin(\pi y),
\end{equation}
where we omit the tilde over the variables for clarity purposes. Due to the nondimensionalization given by Eq.~(\ref{eq:dimless}), the elliptic subproblem given in Eq.~(\ref{eq:poi}) remains the same. In the dimensionless form given in Eq,~(\ref{eq:nbve}), there are only two nondimensional parameters, Reynolds and Rossby numbers, which are related to the physical parameters in the following way:
\begin{equation}\label{eq:ReRo}
Re = \frac{V L}{\nu}, \quad Ro = \frac{V}{\beta L^2}.
\end{equation}
We highlight that the definitions of $Re$ and $Ro$ vary according to the nondimensionalization procedure (see \cite{fox2005reevaluating}, for example).
The following two length scales are also useful for physical understanding of the problems in physical oceanography: (i) the Munk scale, $\delta_M$, for the viscous boundary layer; this is related to small scale dissipation, and (ii) the Rhines scale, $\delta_I$, for the inertial boundary layer; this is measuring the strength of the nonlinearity. These length scales are defined through the following formulas:
\begin{equation}\label{eq:ReRo}
\frac{\delta_M}{L} = \Big(\frac{\nu}{\beta L^3}\Big)^{1/3} = (Re^{-1}Ro)^{1/3}, \quad \quad \frac{\delta_I}{L} = \Big(\frac{V}{\beta L^2}\Big)^{1/2} = (Ro)^{1/2}.
\end{equation}
We note that the specification of these length scales also uniquely determines the $Re$ and $Ro$ numbers in Eq.~(\ref{eq:nbve}). Finally, in order to completely specify the mathematical model, boundary and initial conditions need to be prescribed. In many theoretical studies of large scale ocean circulation models, slip or no-slip boundary conditions are used in simplified Cartesian oceanic basins. Following \cite{cummins1992inertial,ozgokmen1998emergence,greatbatch2000four,holm2003modeling,san2011approximate,san2013approximate},
we use slip boundary conditions for the velocity, which translate into homogeneous Dirichlet boundary conditions for the vorticity: $\omega|_{\Gamma} = 0$, where $\Gamma$ symbolizes all the Cartesian boundaries. The corresponding impermeability boundary condition is imposed as $\psi|_{\Gamma} = 0$. For the initial condition, we start our computations from a quiescent state (i.e., $\omega = 0$, and $\psi=0$) and integrate Eq.~(\ref{eq:nbve}) until we obtain a statistically steady state in which the wind forcing, dissipation, and Jacobian balance each other.

\section{Reduced-order modeling of BVE}
\vspace{-2pt}
\label{sec:rom}
In this section, we develop a POD-ROM for the BVE given by Eq.~(\ref{eq:nbve}). We construct our POD-ROM from the field variable $\omega$ and $\psi$ on the flow domain $\Omega$  at different times, also called snapshots. In this study, the snapshots are obtained by solving Eq.~(\ref{eq:nbve}) using an accurate numerical simulation, which will be briefly described in Section \ref{sec:num}. The main procedure in reduced-order modeling consists of a basis building procedure (i.e., generating POD basis functions) coupled with a model building step (i.e., performing Galerkin projection to obtain the ROM).

\subsection{Computing the POD basis functions}
\label{sec:basis}
In the time marching process of solving Eq.~(\ref{eq:nbve}), the $i$th record of vorticity field variable at time $t=t_i$ is denoted $\omega^{i}(x,y)$ for $i=1,2, ..., N$, where $N$ is the number of snapshots used to build the POD basis. In order to obtain the POD basis functions, we first construct a correlation matrix in the following way:
\begin{equation}\label{eq:cor}
    C_{ij}=\int_{\Omega} \omega^{i} \omega^{j} dx dy,
\end{equation}
where $\Omega$ is the entire spatial domain in which the field variables are defined, and $i$ and $j$ refer to the $i$th and $j$th snapshots. The data correlation matrix $C$ is a non-negative Hermitian matrix. Defining the inner product for two functions $f$ and $g$ as
\begin{equation}\label{eq:inner}
    (f,g)=\int_{\Omega} f g dx dy,
\end{equation}
Eq.~(\ref{eq:cor}) can be written as $C_{ij}=(\omega^{i},\omega^{j})$. Solving the eigenvalue problem for this $C$ matrix provides the optimal POD basis functions. This procedure has been described in detail in the POD literature (e.g., see \cite{sirovich1987turbulence,holmes1998turbulence,ravindran2000reduced}). The eigenvalue problem can be written in the following form:
\begin{equation}\label{eq:eig}
    C\Upsilon=\Upsilon\Lambda,
\end{equation}
where $\Lambda=\mbox{diag}[\lambda_1,\lambda_2, ..., \lambda_N]$, $\Upsilon=[\boldsymbol\upsilon^{1}, \boldsymbol\upsilon^{2},..., \boldsymbol\upsilon^{N}]$, $\lambda_{i}$ is the $i$th eigenvalue, and $\boldsymbol\upsilon^{i}$ refers the corresponding $i$th eigenvector. The $\Upsilon$ matrix is also called right eigenvector matrix; columns are eigenvectors of the correlation matrix $C$.  For practical purposes, the eigenvalues should be stored in descending order, $\lambda_1\geq\lambda_2\geq ...\geq \lambda_N$. The POD basis functions associated with the field variable $\omega$ can be written as
\begin{equation}\label{eq:bas}
    \phi_{1} = \sum_{i=1}^{N} \upsilon_{i}^{1}\omega^{i}, \quad \phi_{2} = \sum_{i=1}^{N} \upsilon_{i}^{2}\omega^{i}, \quad ..., \quad \phi_{N} = \sum_{i=1}^{N} \upsilon_{i}^{N}\omega^{i},
\end{equation}
where $\upsilon_{i}^{j}$ is the $i$th component of eigenvector $\boldsymbol\upsilon^{j}$. The eigenvectors must be normalized in such a way that the basis functions satisfy the following orthogonality condition:
\begin{equation}\label{eq:cond}
    (\phi_{k},\phi_{l})=
    \left\{
      \begin{array}{ll}
    1, & \quad k=l; \\
    0, & \quad k\neq l.
      \end{array}
    \right.
\end{equation}

It can be shown that the eigenvector $\boldsymbol\upsilon^{j}$ must satisfy the following equation for Eq.~(\ref{eq:cond}) to be true \cite{esfahanian2009equation}:
\begin{equation}\label{eq:defi}
    \sum_{i=1}^{N}\upsilon_{i}^{j}\upsilon_{i}^{j}=\frac{1}{\lambda_{j}}.
\end{equation}
In practice, most of the subroutines for solving the eigensystem given in Eq.~(\ref{eq:eig}) return the right eigenvector matrix $\Upsilon$ such that all the eigenvectors are normalized to unity. In that case, the orthogonal POD basis functions can be obtained as
\begin{equation}\label{eq:PODbasis}
    \phi_{j}(x,y) =\frac{1}{\lambda_{j}} \sum_{i=1}^{N} \upsilon_{i}^{j}\omega^{i}(x,y),
\end{equation}
where $\phi_{j}(x,y)$ is the $j$th POD basis function of the corresponding field $\omega(x,y)$.

\subsection{Galerkin projection to obtain ROM}
\label{sec:Galerkin}
DNS computations of Eq.~(\ref{eq:nbve}) provide snapshots at different time steps. The correlation matrix for the vorticity field is then generated using $N$ snapshots (i.e., from the fields at different time instances $t_1$, $t_2$, ..., $t_N$). First, these data sets are decomposed into the mean part and the fluctuating components:
\begin{equation}\label{eq:decom1}
    \omega(x,y,t_i) = \bar{\omega}(x,y) + \omega'(x,y,t_i), \quad \bar{\omega}(x,y) = \frac{1}{N}\sum_{i=1}^{N} \omega(x,y,t_i),
\end{equation}
where $\bar{\omega}$ is the mean part which is function of only the space variables, and $\omega'$ is the fluctuating part, which is function of both space variables and time. The mean stream function, $\bar{\psi}$, can be defined in the same way. Then, the correlation matrix $C$ is obtained from the data set of the fluctuating part. After solving the eigensystem for the set from the vorticity field, the corresponding POD basis functions become
\begin{equation}\label{eq:basisw}
    \phi_{j}(x,y) =\frac{1}{\lambda_{j}} \sum_{i=1}^{N} \upsilon_{i}^{j}\omega'(x,y,t_i),
\end{equation}
where $\phi_{j}$ is the $j$th POD basis function of the vorticity field. Here, $\omega'(x,y,t_i)$ represents the fluctuating components of the $i$th snapshot of the vorticity field, $\lambda_{j}$ is the $j$th eigenvalues of the vorticity field, and $\upsilon_{i}^{j}$ is the $i$th components of the corresponding eigenvector for the vorticity field. Using the the kinematic relationship between stream function and vorticity given by Eq.~(\ref{eq:poi}), the $j$th basis function for the stream function, $\varphi_{j}(x,y)$, can be obtained from the $j$th vorticity basis function by solving a Poisson equation
\begin{equation}\label{eq:psi}
    \nabla^2 \varphi_{j} = -\phi_{j}.
\end{equation}

These basis functions account for the essential dynamics of the system. To generate a ROM, we truncate the system by considering the first $R$ basis functions out of the total $N$ bases, where $R\ll N$. These largest energy containing $R$ modes correspond to the $R$ largest eigenvalues, $\lambda_1$, $\lambda_2$, ..., $\lambda_R$. In the POD-ROM framework, the field variables can be constructed by using these basis functions in the following way:
\begin{equation}\label{eq:const1}
    \omega'(x,y,t) =  \sum_{k=1}^{R} \alpha_{k}(t)\phi_{j}(x,y),
\end{equation}
where we decompose $\omega'(x,y,t)$ using time dependent coefficient $\alpha_k$ and the space dependent modes $\phi_{j}$.
We emphasize that the same $\alpha_k$ are defined for both the vorticity and stream function fields.

To obtain a ROM, we rearrange Eq.~(\ref{eq:nbve}) using linear and nonlinear operators in the following form:
\begin{equation}\label{eq:vsa1}
\frac{\partial \omega}{\partial t} = N[\omega; \psi] + L[\omega] + H[\psi] + F,
\end{equation}
where, for arbitrary functions $f$ and $g$, the linear operators $L$ and $H$ and the nonlinear operator $N$ are given by
\begin{equation}\label{eq:ope}
    L[f] =  \frac{1}{Re}\Big(\frac{\partial^2 f }{\partial x^2} + \frac{\partial^2 f}{\partial y^2}\Big), \quad H[f] = \frac{1}{Ro}\frac{\partial f}{\partial x}, \quad N[f;g]= -\frac{\partial g}{\partial y}\frac{\partial f}{\partial x} + \frac{\partial g}{\partial x}\frac{\partial f}{\partial y}.
\end{equation}

Next, we apply the Galerkin projection by multiplying Eq.~(\ref{eq:vsa1}) with the basis functions and integrating over the domain $\Omega$. Using the inner product definition given by Eq.~(\ref{eq:inner}), the Galerkin projection on the $k$th basis function can be written as
\begin{equation} \label{eq:vsb1}
  \Big(\frac{\partial \omega}{\partial t},\phi_{k}\Big)  = \Big(N[\omega; \psi], \phi_{k} \Big) +\Big(L[\omega],\phi_{k}\Big) + \Big( H[\psi], \phi_{k} \Big) + \Big(\frac{1}{Ro}\sin(\pi y),\phi_{k} \Big).
\end{equation}

Substituting Eq.~(\ref{eq:decom1}) and Eq.~(\ref{eq:const1}) into Eq.~(\ref{eq:vsb1}), and simplifying the resulting equations by using the orthogonality condition given by Eq.~(\ref{eq:cond}), we obtain the following coupled POD reduced-order system for $k=1,2, ..., R$:
\begin{equation} \label{eq:rom1}
  \frac{d \alpha_k}{d t} = B_{k} + \sum_{i=1}^{R} P_{ik}\alpha_{i} + \sum_{i=1}^{R}\sum_{j=1}^{R} Q_{ijk}\alpha_{i}\alpha_{j},
\end{equation}
where
\begin{eqnarray}
  & & B_{k} = \big( L[\bar{\omega}], \phi_{k} \big) +  \big( H[\bar{\psi}], \phi_{k} \big) + \big( N[\bar{\omega};\bar{\psi}], \phi_{k} \big) + \Big(\frac{1}{Ro}\sin(\pi y),\phi_{k} \Big)  , \label{eq:roma1}\\
  & & P_{ik} = \big( L[\phi_{i}], \phi_{k} \big) +\big( H[\varphi_{i}], \phi_{k} \big) + \big( N[\bar{\omega};\varphi_{i}] +  N[\phi_{i};\bar{\psi}] , \phi_{k} \big)  , \label{eq:roma3}\\
  & & Q_{ijk} =  \big( N[\phi_{i};\varphi_{j}], \phi_{k} \big). \label{eq:roma7}
\end{eqnarray}

The POD-ROM given by Eq.~(\ref{eq:rom1}) consists of $R$ coupled ordinary differential equations and can be easily solved by a standard numerical method (a third-order Runge-Kutta scheme is used in this paper). We emphasize that the number of degrees of freedom of the system has been substantially decreased and the resulting dynamical system can be solved very efficiently, since all the POD basis functions and corresponding model coefficients given by Eqs.~(\ref{eq:roma1})-(\ref{eq:roma7}) are precomputed from the data provided by snapshots. %The major criticism for POD methodology is that the procedure needs a set of snapshots to generate basis functions.
The POD basis functions are usually obtained from a fine level computation, such as a DNS.
%performed by a fourth-order Arakawa scheme in this study.
%In that sense, POD can be considered a post processing methodology, rather than a physical system solver, but could be very viable tool for real time analysis for parameter identification, optimization, and control applications. In this study, however, we focus on the predictive capabilities of various POD reduced-order models for the wind-driven ocean circulation problems in quasigeostrophic system.
To complete the dynamical system given by Eq.~(\ref{eq:rom1}), the initial condition is specified by using the following projection
\begin{equation}\label{eq:ic1}
\alpha_{k}(t_{in}) = \Big( \omega(x,y,t_{in}) - \bar{\omega}(x,y), \phi_{k} \Big),
\end{equation}
where $\omega(x,y,t_{in})$ is the vorticity field specified at the initial time $t_{in}$.

The POD-ROM given by Eq.~(\ref{eq:rom1}) usually works well for a relatively smooth system for which the largest $R$ modes adequately capture the system's dynamics. One of the main sources of inaccuracy in the POD-ROM is the truncation of the higher-order modes. Stabilization schemes often improve the performance of the POD-ROM \cite{kalb2007intrinsic,bergmann2009enablers,wang2012proper}. The first and simplest model to overcome errors due to the finite truncation involved in the POD-ROM approach for complex systems is called Heisenberg stabilization and uses a global constant eddy viscosity coefficient \cite{bergmann2009enablers}. In large eddy simulations of turbulent flows, this stabilization approach is also called mixing length closure; this terminology is also used in POD literature \cite{wang2012proper}. This stabilization model accounts for the effects of the truncated modes by introducing a constant eddy viscosity coefficient in the model. Therefore, the corresponding physical physical parameter in the dissipation mechanism, which is $Re$ in our system, can be modified by adding an eddy viscosity coefficient in the following form:
\begin{equation}\label{eq:heis}
    \frac{1}{Re}\Rightarrow \frac{1}{Re}(1+\nu_{a}),
\end{equation}
where the free stabilization parameter $\nu_a$ is considered as a global constant for all the modes in this model (i.e., $\nu_e = \nu_a/Re$ is the total amount of eddy viscosity added to the system). The constant eddy viscosity idea suggested in Eq.~(\ref{eq:heis}) can be improved by supposing that the amount of dissipation is not identical for all the POD modes \cite{rempfer1991,cazemier1997}. In our study, the global viscosity is replaced by modal viscosities using a linear kernel in the following form:
\begin{equation}\label{eq:rem}
    \frac{1}{Re}\Rightarrow \frac{1}{Re}(1+\nu_{a}\frac{k}{R}),
\end{equation}
where the constant $\nu_a$ is now defined as the amplitude of the eddy viscosity stabilization. Using a linear viscosity kernel, $k/R$, the amount of dissipation and hence stabilization increases linearly with the POD modal index $k$. Thus, we add a small amount of dissipation to the ROM for the smaller POD index representing more energy content in the system.

The problem is then to adjust the constant $\nu_a$ in order to obtain a better accuracy in the POD-ROM. An important aspect of this eddy viscosity stabilization model is that it does not require any additional computational cost for computing the ROM coefficient in Eqs.~(\ref{eq:roma1})-(\ref{eq:roma7}). Therefore, a sensitivity analysis with respect to the free parameter $\nu_a$ has a computational cost similar to the cost of solving the ROM given in Eq.~(\ref{eq:rom1}). Thus, the optimal value of the $\nu_a$ coefficient can be obtained efficiently. We also emphasize that specifying $\nu_a=0$ yields the standard Galerkin POD-ROM.

\section{Numerical schemes}
\vspace{-2pt}
\label{sec:num}
%The objective of the present work is to develop an efficient proper orthogonal decomposition reduced-order model for BVE equation.
In this section, we provide a brief description of the numerical methods employed in this study.

\subsection{The fourth-order Arakawa scheme}
\vspace{-2pt}
\label{sec:ara}
Arakawa \cite{arakawa1966computational} suggested that the conservation of energy, enstrophy and skew symmetry is sufficient to avoid computational instabilities arising from nonlinear interactions. The conservation and stability properties of the Arakawa scheme were investigated by Lilly \cite{lilly1965computational} by means of spectral analysis along with several first and second order time integration methods. Using the third- and fourth-order Runge-Kutta methods, this scheme was also tested to compute decaying two-dimensional turbulence simulations \cite{san2012high}. The nonlinear convective terms in Eq.~(\ref{eq:nbve}) were defined as the Jacobian
\begin{equation}
J(\omega,\psi) = \frac{\partial \psi}{\partial y}\frac{\partial \omega}{\partial x} - \frac{\partial \psi}{\partial x}\frac{\partial \omega}{\partial y}.
\label{eq:jac}
\end{equation}
The second-order Arakawa scheme for the Jacobian is
\begin{equation}
J_{I}(\omega,\psi) =\frac{1}{3}\Big(J_{1}(\omega,\psi)+J_{2}(\omega,\psi)+J_{3}(\omega,\psi)\Big)
\label{eq:ja1}
\end{equation}
where the discrete parts of the Jacobian are
\begin{eqnarray}
J_{1}(\omega,\psi) &=& \frac{1}{4h_xh_y}\Big[(\omega_{i+1,j}-\omega_{i-1,j})(\psi_{i,j+1}-\psi_{i,j-1}) \nonumber \\
&-&(\omega_{i,j+1}-\omega_{i,j-1})(\psi_{i+1,j}-\psi_{i-1,j})\Big]
\label{eq:j1}
\end{eqnarray}
\begin{eqnarray}
J_{2}(\omega,\psi) &=& \frac{1}{4h_xh_y}\Big[\omega_{i+1,j}(\psi_{i+1,j+1}-\psi_{i+1,j-1}) -\omega_{i-1,j}(\psi_{i-1,j+1}-\psi_{i-1,j-1}) \nonumber \\
&-&\omega_{i,j+1}(\psi_{i+1,j+1}-\psi_{i-1,j+1}) +\omega_{i,j-1}(\psi_{i+1,j-1}-\psi_{i-1,j-1}) \Big]
\label{eq:j2}
\end{eqnarray}
\begin{eqnarray}
J_{3}(\omega,\psi) &=& \frac{1}{4h_xh_y}\Big[\omega_{i+1,j+1}(\psi_{i,j+1}-\psi_{i+1,j}) -\omega_{i-1,j-1}(\psi_{i-1,j}-\psi_{i,j-1}) \nonumber \\
&-&\omega_{i-1,j+1}(\psi_{i,j+1}-\psi_{i-1,j}) +\omega_{i+1,j-1}(\psi_{i+1,j}-\psi_{i,j-1}) \Big].
\label{eq:j3}
\end{eqnarray}
The fourth-order accurate Arakawa discretization of the Jacobian becomes
\begin{equation}
J_{II}(\omega,\psi)  =\frac{1}{3}\Big(J_{4}(\omega,\psi)+J_{5}(\omega,\psi)+J_{6}(\omega,\psi)\Big),
\label{eq:ja2}
\end{equation}
where
\begin{eqnarray}
J_{4}(\omega,\psi) &=& \frac{1}{8h_xh_y}\Big[(\omega_{i+1,j+1}-\omega_{i-1,j-1})(\psi_{i-1,j+1}-\psi_{i+1,j-1}) \nonumber \\
&-&(\omega_{i-1,j+1}-\omega_{i+1,j-1})(\psi_{i+1,j+1}-\psi_{i-1,j-1})\Big]
\label{eq:j4}
\end{eqnarray}
\begin{eqnarray}
J_{5}(\omega,\psi) &=& \frac{1}{8h_xh_y}\Big[\omega_{i+1,j+1}(\psi_{i,j+2}-\psi_{i+2,j}) -\omega_{i-1,j-1}(\psi_{i-2,j}-\psi_{i,j-2}) \nonumber \\
&-&\omega_{i-1,j+1}(\psi_{i,j+2}-\psi_{i-2,j}) +\omega_{i+1,j-1}(\psi_{i+2,j}-\psi_{i,j-2}) \Big]
\label{eq:j5}
\end{eqnarray}
\begin{eqnarray}
J_{6}(\omega,\psi) &=& \frac{1}{8h_xh_y}\Big[\omega_{i+2,j}(\psi_{i+1,j+1}-\psi_{i+1,j-1}) -\omega_{i-2,j}(\psi_{i-1,j+1}-\psi_{i-1,j-1}) \nonumber \\
&-&\omega_{i,j+2}(\psi_{i+1,j+1}-\psi_{i-1,j+1}) +\omega_{i,j-2}(\psi_{i+1,j-1}-\psi_{i-1,j-1}) \Big].
\label{eq:j6}
\end{eqnarray}
Arakawa showed that $J_{II}$ conserves enstrophy and energy and the following Jacobian
\begin{equation}
J(\omega,\psi) = 2J_{I}(\omega,\psi)  - J_{II}(\omega,\psi)  + O(h^{4})
\label{eq:ja2}
\end{equation}
has fourth-order accuracy. The linear terms (i.e., the rotation and viscous dissipation terms) can be discretized with the fourth-order explicit difference scheme. For any scalar value $f$, the classical centered difference schemes for the second-derivative up to the fourth-order accuracy are given by \cite{strikwerda2007finite}
\begin{eqnarray}\label{eq:ecd}
    f_{i}^{''} &=& \frac{1}{h^2} (f_{i-1} -2f_{i} + f_{i+1}) +O(h^2), \\
    f_{i}^{''} &=& \frac{1}{12h^2} (-f_{i-2}+16f_{i-1} -30f_{i} + 16f_{i+1} - f_{i+2}) +O(h^4),
\end{eqnarray}
where $h$ is the step size in the derivative direction.
In vorticity-steam function formulation, Briley's formula is usually used to satisfy the no-slip boundary condition at the walls. In this formula the vorticity values at the boundary are computed from the stream function values according to the following third-order accurate formula \cite{briley1971numerical}:
\begin{equation}\label{eq:cd2}
    \omega_{0} = \frac{1}{h^2}\Big( \frac{85}{18} \psi_{0} - 6 \psi_{1} +\frac{3}{2}\psi_{2} - \frac{2}{9}\psi_{3}\Big) ,
\end{equation}
where the subscript $0$ represents the discrete point index on the free-slip boundary where we set $\psi_{0}=0$.

\subsection{Time advancement scheme}
\label{sec:rk}
Semi-discrete ordinary differential equations are obtained after a spatial discretization of the partial differential equations. To implement the Runge-Kutta scheme for the time integration, we cast the model equations in the following form
\begin{equation}\label{eq:cast}
    \frac{d\omega}{dt} = \pounds (\omega;\psi) ,
\end{equation}
where $\pounds (\omega;\psi)$ is the discrete operator of spatial derivatives including nonlinear convective terms and linear rotational and diffusive terms. It should be also noticed that the set of ordinary differential equations for the amplitudes $\alpha_{k}$ of POD-ROM can also be written in a similar form. In both DNS and ROM we assume that the numerical approximation for time level $n$ is known, and we seek the numerical approximation for time level $n+1$, after the time step $\Delta t$. The optimal third-order accurate total variation diminishing Runge-Kutta (TVDRK3) scheme is then given as \cite{gottlieb1998total}
\begin{eqnarray}
\omega^{(1)} &=& \omega^{n} + \Delta t \pounds(\omega^{n};\psi^{n}), \nonumber \\
\omega^{(2)} &=& \frac{3}{4} \omega^{n} + \frac{1}{4} \omega^{(1)} + \frac{1}{4}\Delta t \pounds (\omega^{(1)};\psi^{(1)}), \nonumber \\
\omega^{n+1} &=& \frac{1}{3} \omega^{n} + \frac{2}{3} \omega^{(2)} + \frac{2}{3}\Delta t \pounds (\omega^{(2)};\psi^{(2)}).
\label{eq:TVDRK}
\end{eqnarray}

\subsection{Numerical integration}
\label{sec:ni}
In order to perform inner products given by Eq.~(\ref{eq:inner}), we compute the integral of $u(x,y)$ over the domain $\Omega$ by using the dual integration method with Simpson's 3/8 rule \cite{hoffman2001numerical}
\begin{equation}\label{eq:ni1}
    \int_{\Omega} u(x,y)dxdy = \frac{1}{3\Delta y} \sum_{j=1}^{N_y/2-1} \big( f_{i,2j} +4f_{i,2j+1} +f_{i,2j+2}   \big),
\end{equation}
where
\begin{equation}\label{eq:ni2}
    f_{i,j}= \frac{1}{3\Delta x} \sum_{i=1}^{N_x/2-1} \big( u_{2i,j} +4f_{2i+1,j} +f_{2i+2,j}   \big)
\end{equation}
and $N_x$ and $N_y$ are even numbers representing the total number of grid points in the $x$ and $y$ directions.

\subsection{Fast Poisson solver}
\label{sec:fps}
The elliptic equations given in Eq.~(\ref{eq:poi}) and Eq.~(\ref{eq:psi}) can be written in the form of $\nabla^2 u =f$. The compact fourth-order discretization scheme with nine point stencil can be written as \cite{wang2009sixth}
\begin{eqnarray}
a u_{i,j} &+& b(u_{i+1,j}+u_{i-1,j}) + c(u_{i,j+1}+u_{i,j-1}) \nonumber \\
          &+& d(u_{i+1,j+1}+u_{i+1,j-1}+u_{i-1,j+1}+u_{i-1,j-1}) \nonumber \\
          &=& e(8f_{i,j}+f_{i+1,j}+f_{i-1,j}+f_{i,j+1}+f_{i,j-1}),
\label{eq:mehr}
\end{eqnarray}
where the coefficients are $a=-10(1+\gamma^2)/2$, $b=5-\gamma^2$, $c=5\gamma^2 - 1$, $d=(1+\gamma^2)/2$, $e=\Delta x^2 / 2$, and $\gamma$ is defined as the mesh aspect ratio, $\gamma=\Delta x/ \Delta y$. For $\gamma=1$ the scheme is well known and is sometimes called Mehrstellen scheme. The no-slip boundary condition implies impermeability condition for the stream function. Therefore, the prescribed boundary condition for the stream function $\psi |_{\Gamma} = 0$ results in homogeneous Drichlet boundary condition (i.e., $u |_{\Gamma} = 0$) on the boundary $\Gamma$ and suggests the use of a fast sine transform. The procedure involves three steps. First, an inverse sine transform for the source term is given by
\begin{equation}\label{eq:isine}
    \hat{f}_{k,l}= \frac{2}{N_x}\frac{2}{N_y}\sum_{i=1}^{N_x-1}\sum_{j=1}^{N_y-1} f_{i,j} \sin\Big(\frac{\pi k i}{N_x} \Big) \sin\Big(\frac{\pi l j}{N_y} \Big),
\end{equation}
where $k$ and $l$ are wavenumbers in Fourier space. Here, the symbol hat is used to represent the corresponding Fourier coefficient of the physical grid valued data with a subscript pair $i$, $j$, where $i=0,1,2, ..., N_x$, and $j=0,1,2, ..., N_y$. As a second step, we directly solve the subproblem in Fourier space using the following relationship:
\begin{equation}\label{eq:solver}
    \hat{u}_{k,l}= \frac{e \hat{f}_{k,l} \Big( 8 + 2\cos\big(\frac{\pi k}{N_x}\big) + 2\cos\big(\frac{\pi l}{N_y}\big)\Big) }{a + 2b\cos\big(\frac{\pi k}{N_x}\big) + 2c\cos\big(\frac{\pi l}{N_y}\big) + 4d\cos\big(\frac{\pi k}{N_x}\big)\cos\big(\frac{\pi l}{N_y}\big)}.
\end{equation}
Finally, the physical values for $u$ are found by performing a forward sine transform:
\begin{equation}\label{eq:fsine}
    u_{i,j}= \sum_{k=1}^{N_x-1}\sum_{l=1}^{N_y-1} \hat{u}_{k,l} \sin\Big(\frac{\pi k i}{N_x} \Big) \sin\Big(\frac{\pi l j}{N_y} \Big).
\end{equation}
This elliptic solver has a computational cost of $O\big( N_x N_y \log(N_x)\log(N_y)\big)$ and is considered as one of the optimal Poisson solvers for Cartesian grid applications (see \cite{san2013coarse} for details). The fast Fourier transform (FFT) algorithm given by Press et al. \cite{press1992numerical} is used for forward and inverse sine transforms.

\section{Results}
\vspace{-2pt}
\label{sec:result}

\begin{figure}
\includegraphics[width=1.0\textwidth]{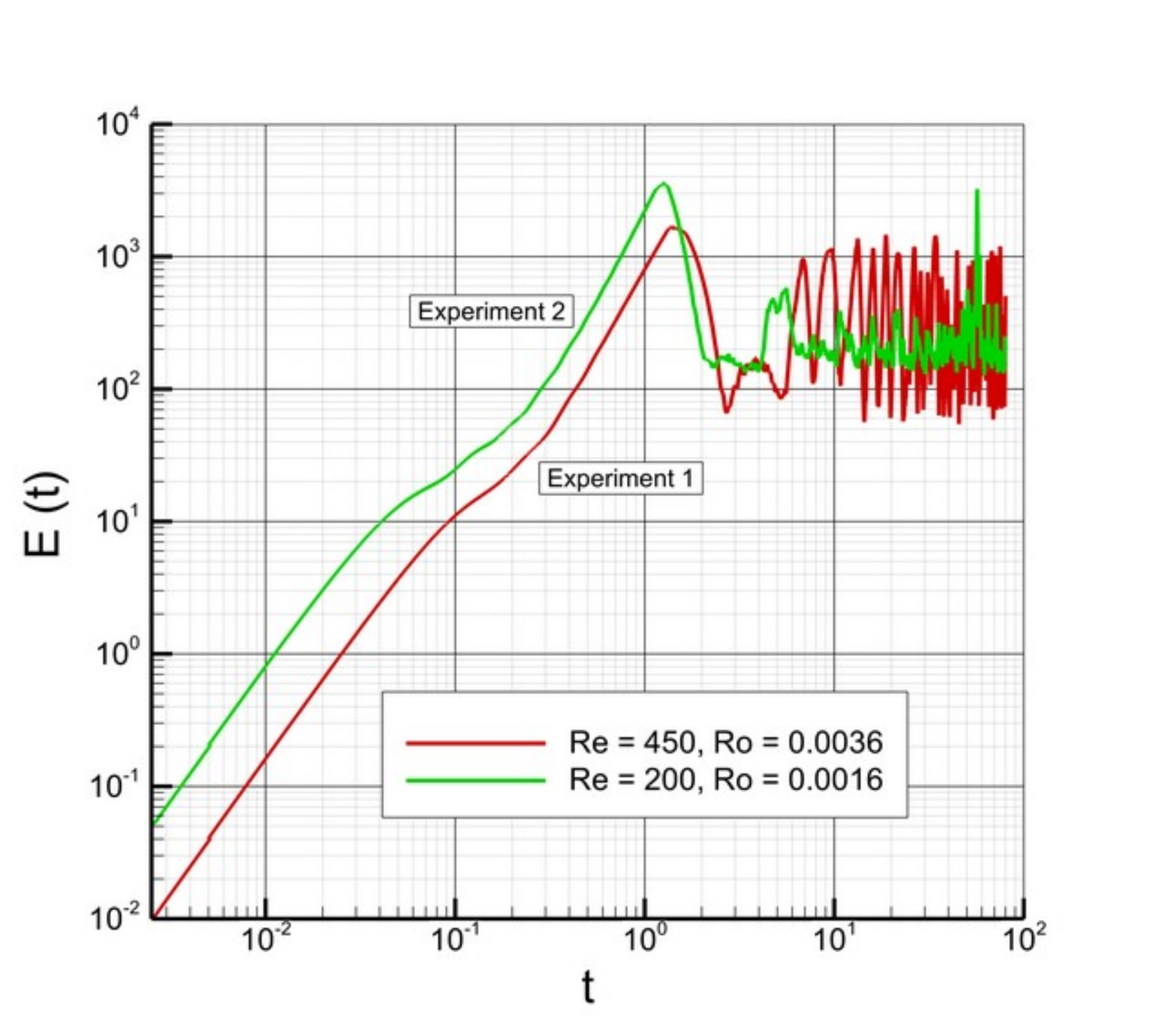}
\caption{Time histories of basin integrated total kinetic energy.}
\label{fig:hist}
\end{figure}

\begin{figure}
\mbox{
\subfigure[$\omega$ ($t=40$)]{\includegraphics[width=0.33\textwidth]{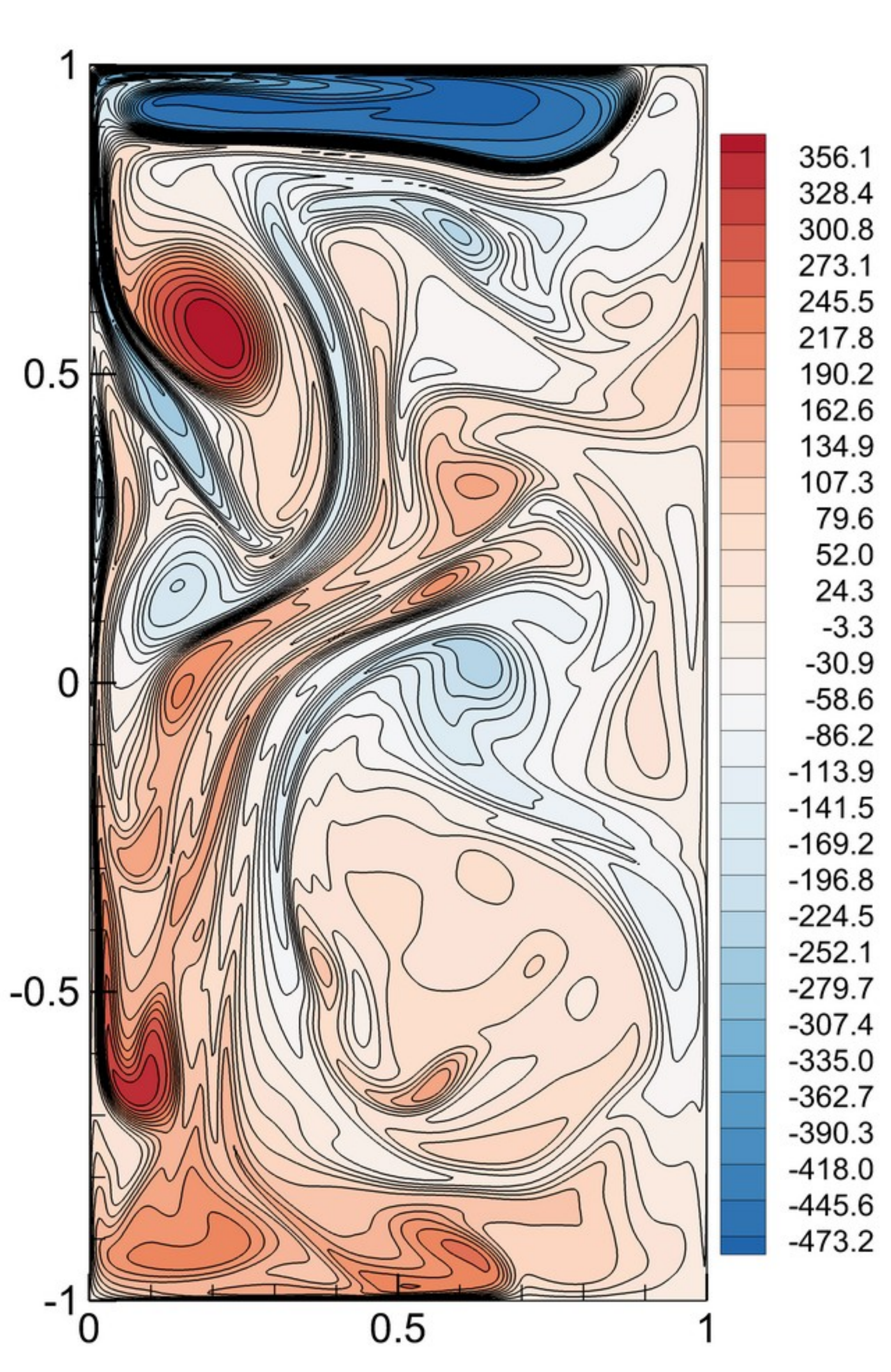}}
\subfigure[$\omega$ ($t=60$)]{\includegraphics[width=0.33\textwidth]{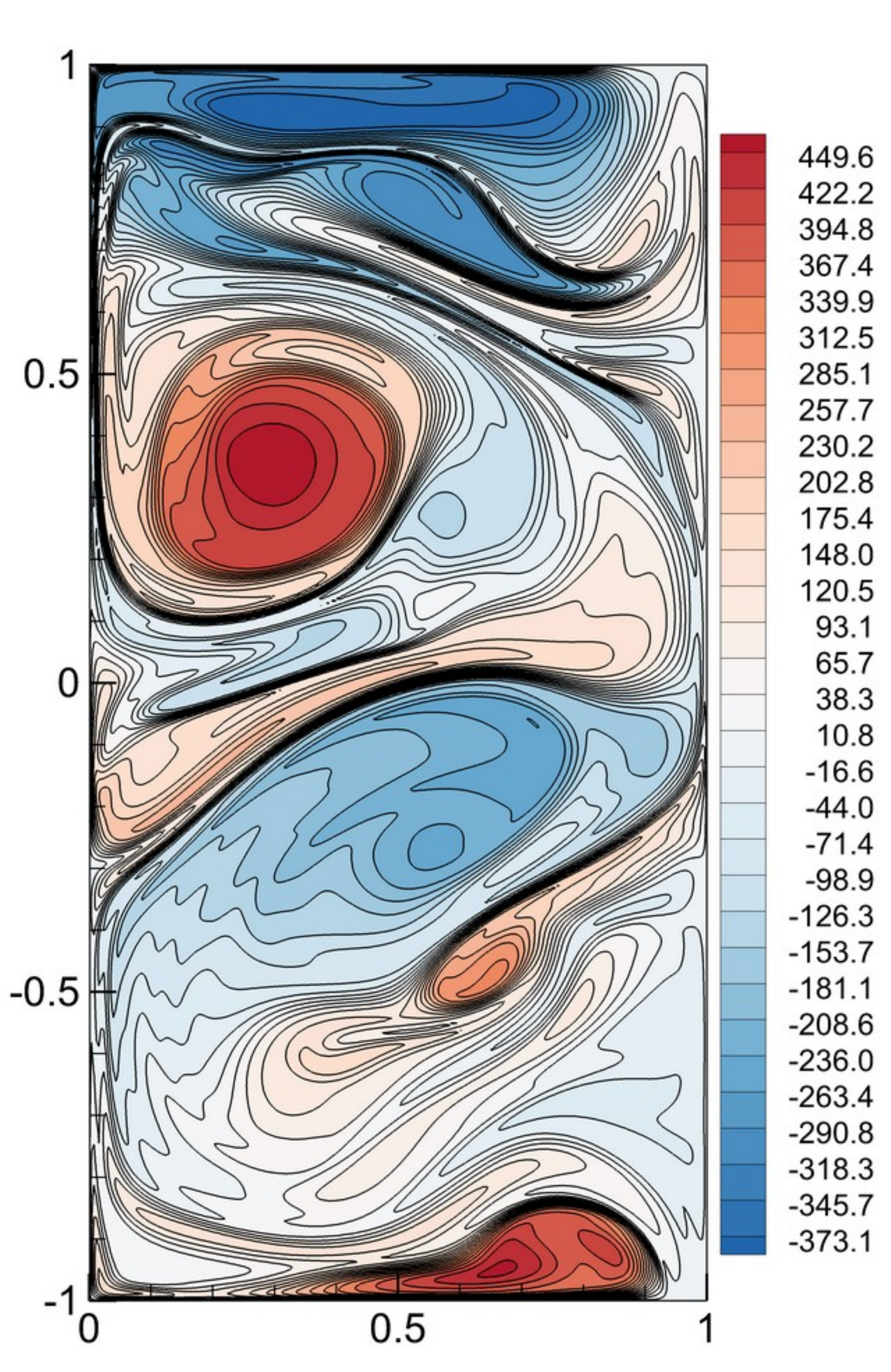}}
\subfigure[$\omega$ ($t=80$)]{\includegraphics[width=0.33\textwidth]{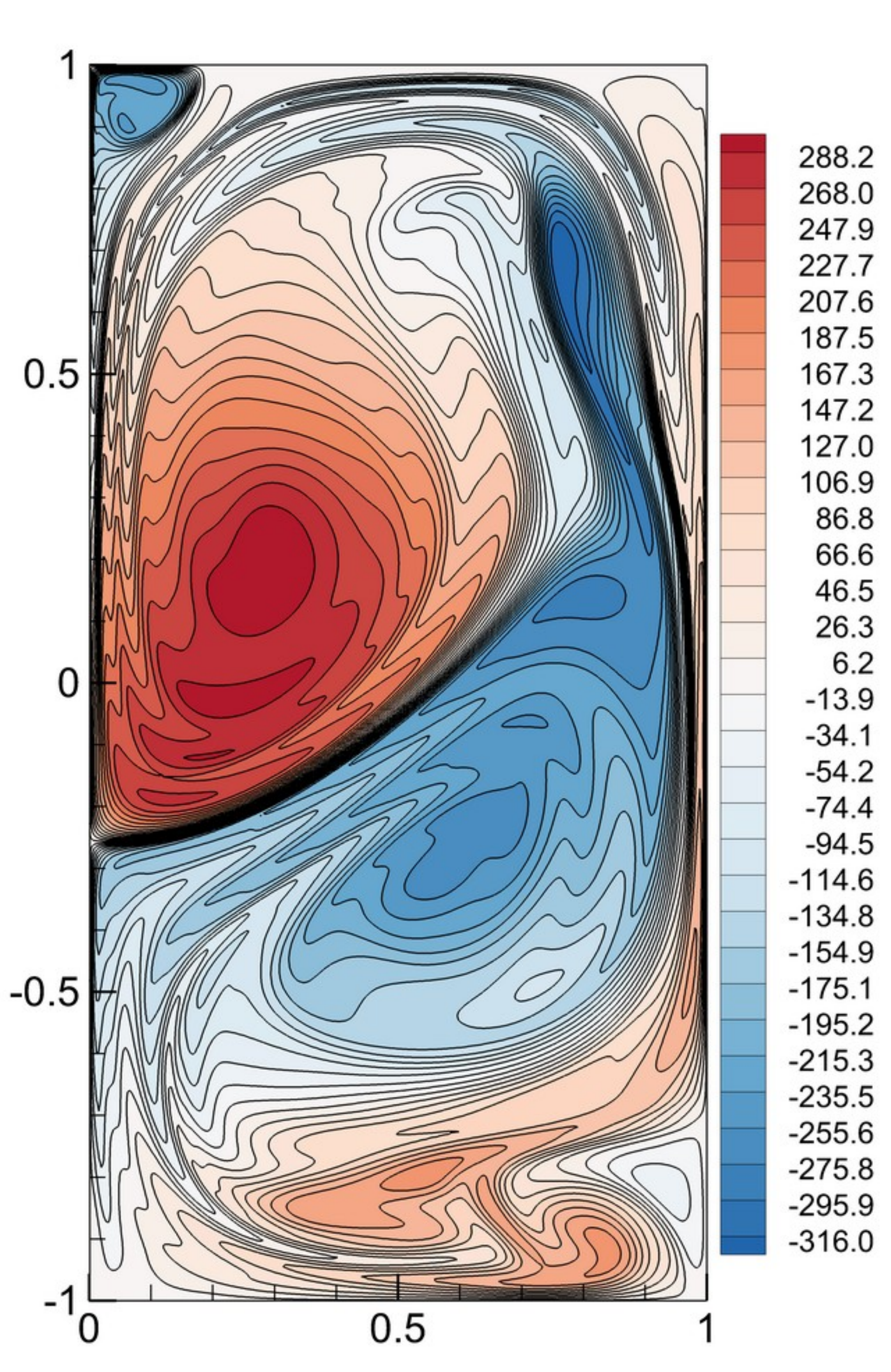}}
}\\
\mbox{
\subfigure[$\psi$ ($t=40$)]{\includegraphics[width=0.33\textwidth]{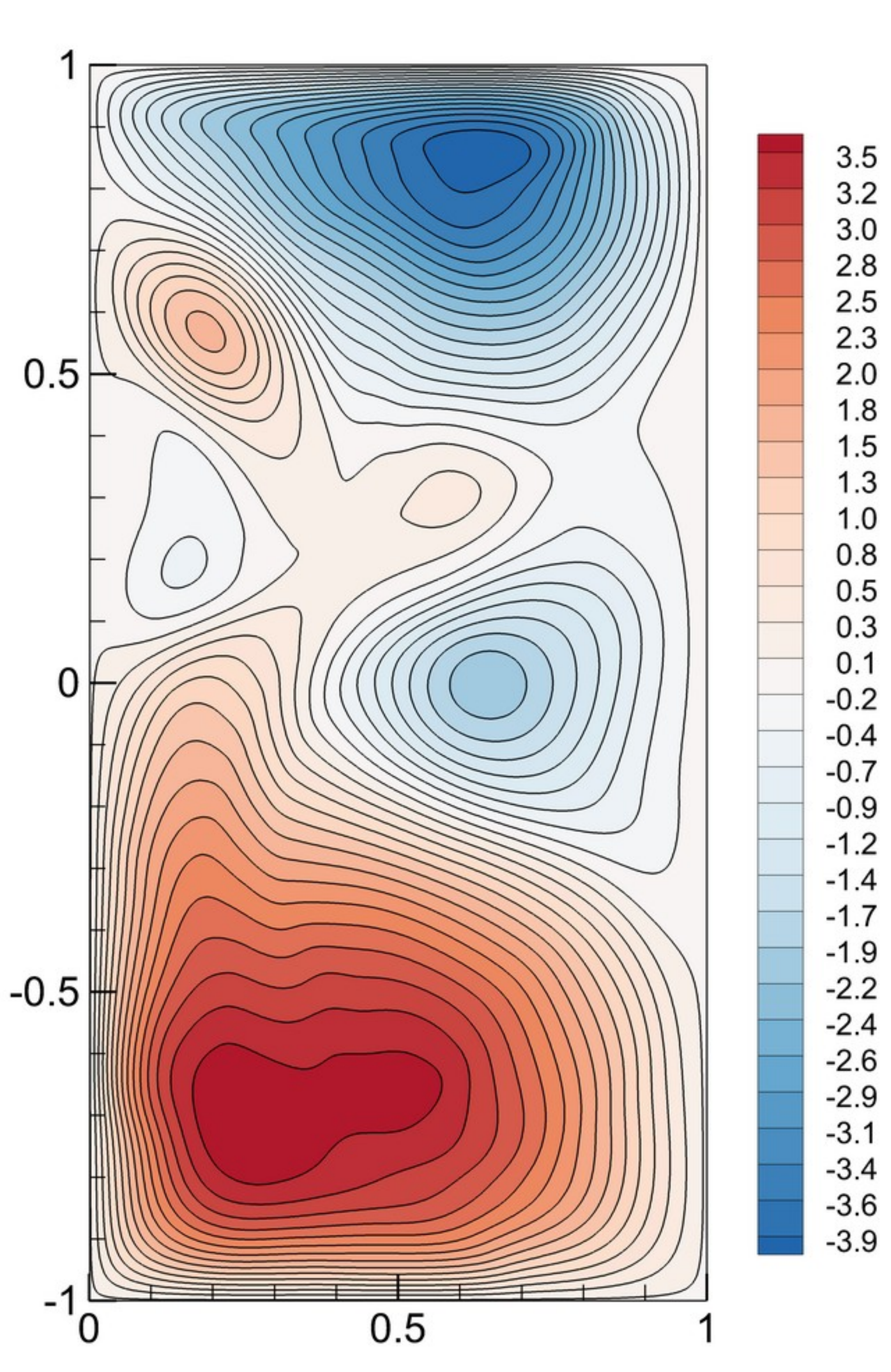}}
\subfigure[$\psi$ ($t=60$)]{\includegraphics[width=0.33\textwidth]{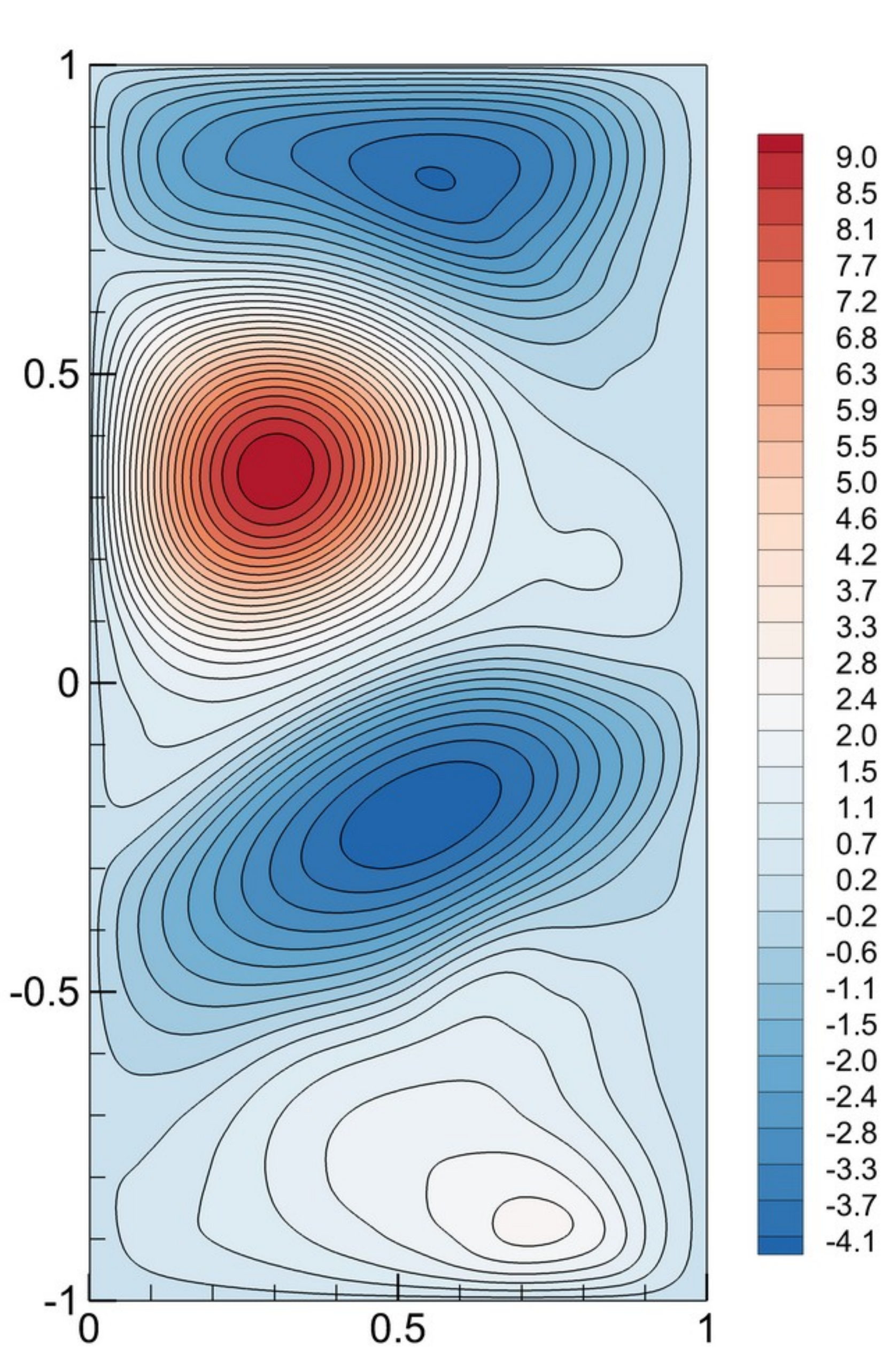}}
\subfigure[$\psi$ ($t=80$)]{\includegraphics[width=0.33\textwidth]{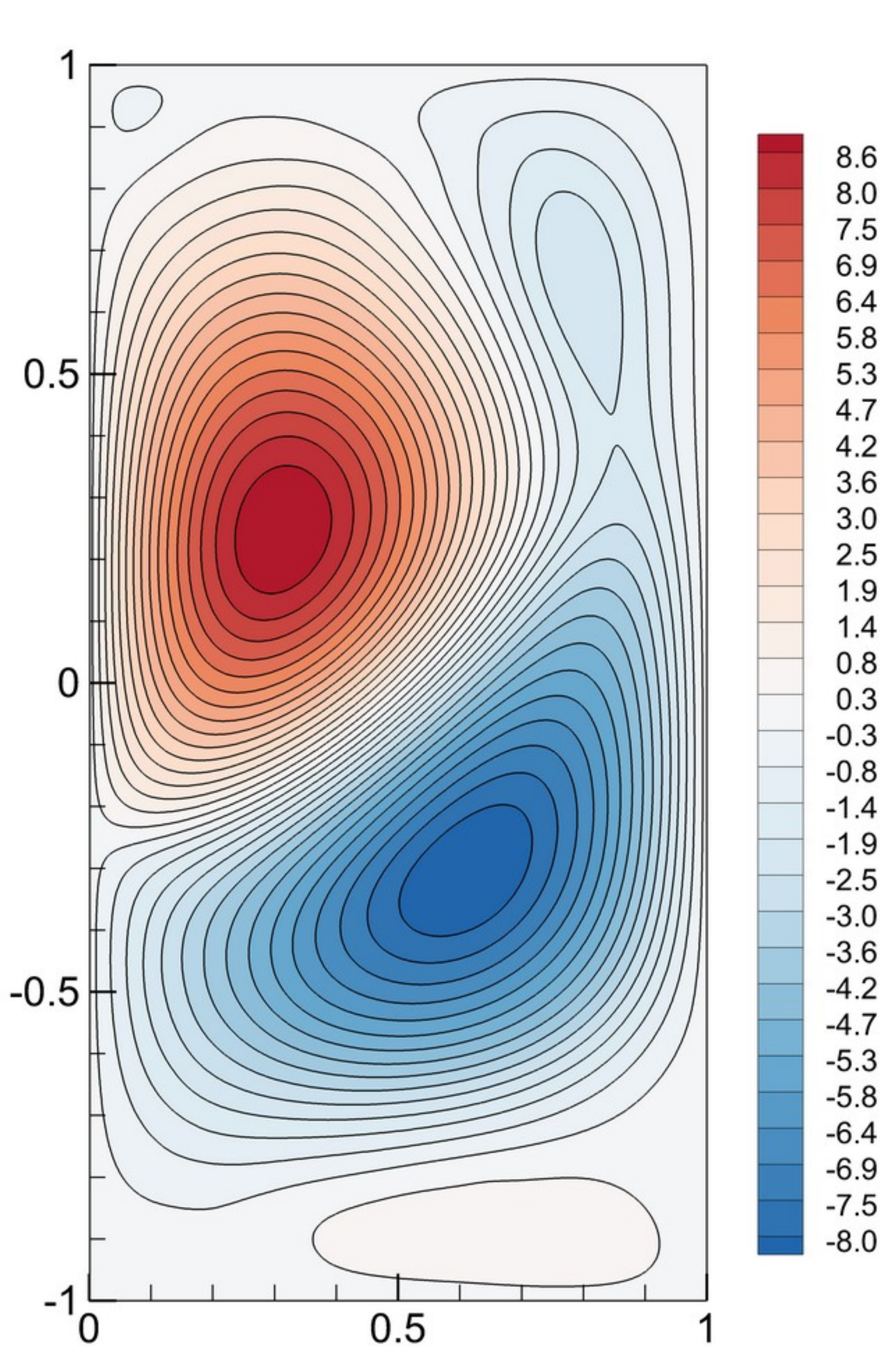}}
}
\caption{Instantaneous vorticity (a-c) and stream function (d-f) contour plots for Experiment 1.}
\label{fig:e1-ins}
\end{figure}

\begin{figure}
\mbox{
\subfigure[$\omega$ ($t=40$)]{\includegraphics[width=0.33\textwidth]{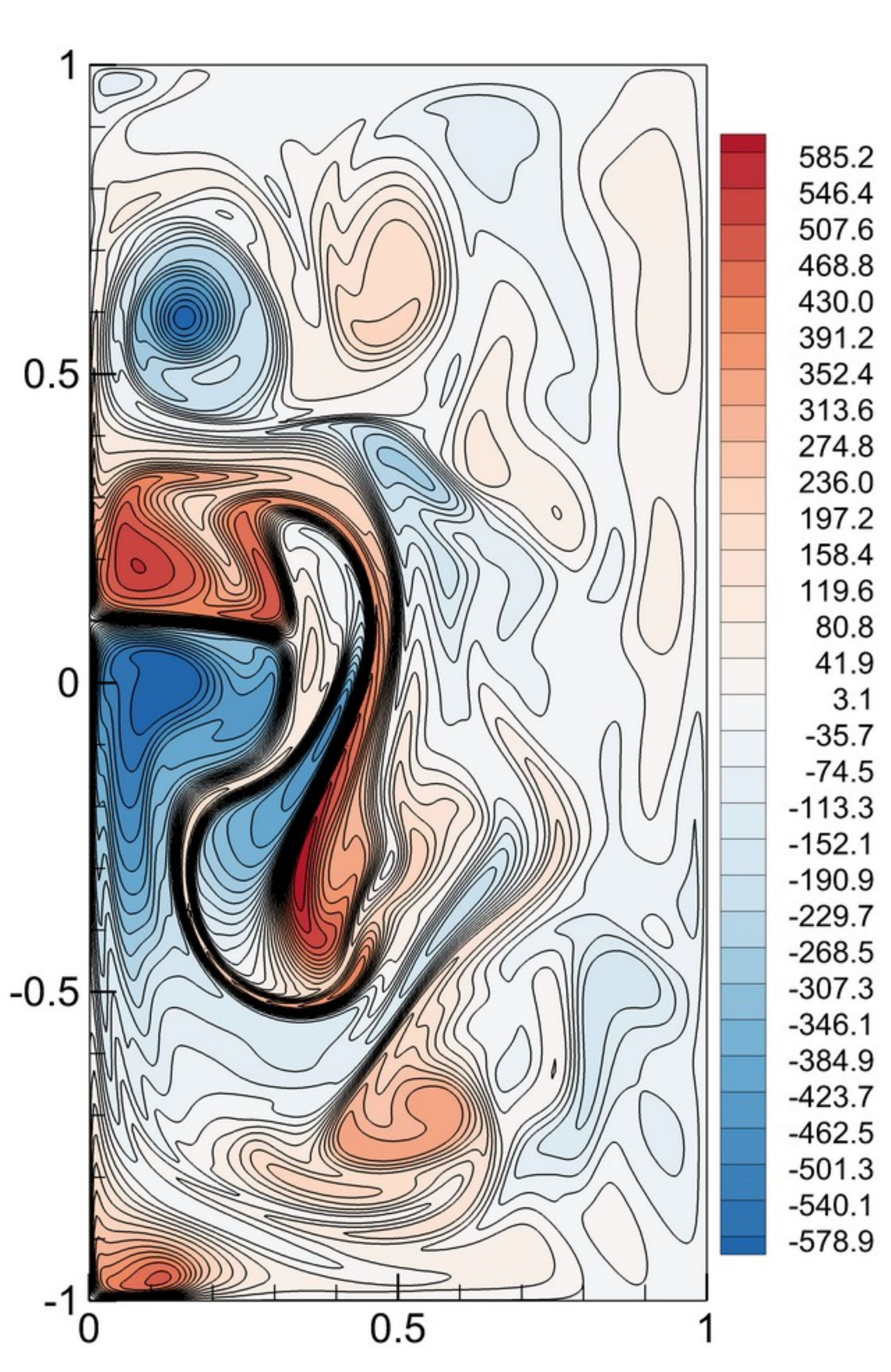}}
\subfigure[$\omega$ ($t=60$)]{\includegraphics[width=0.33\textwidth]{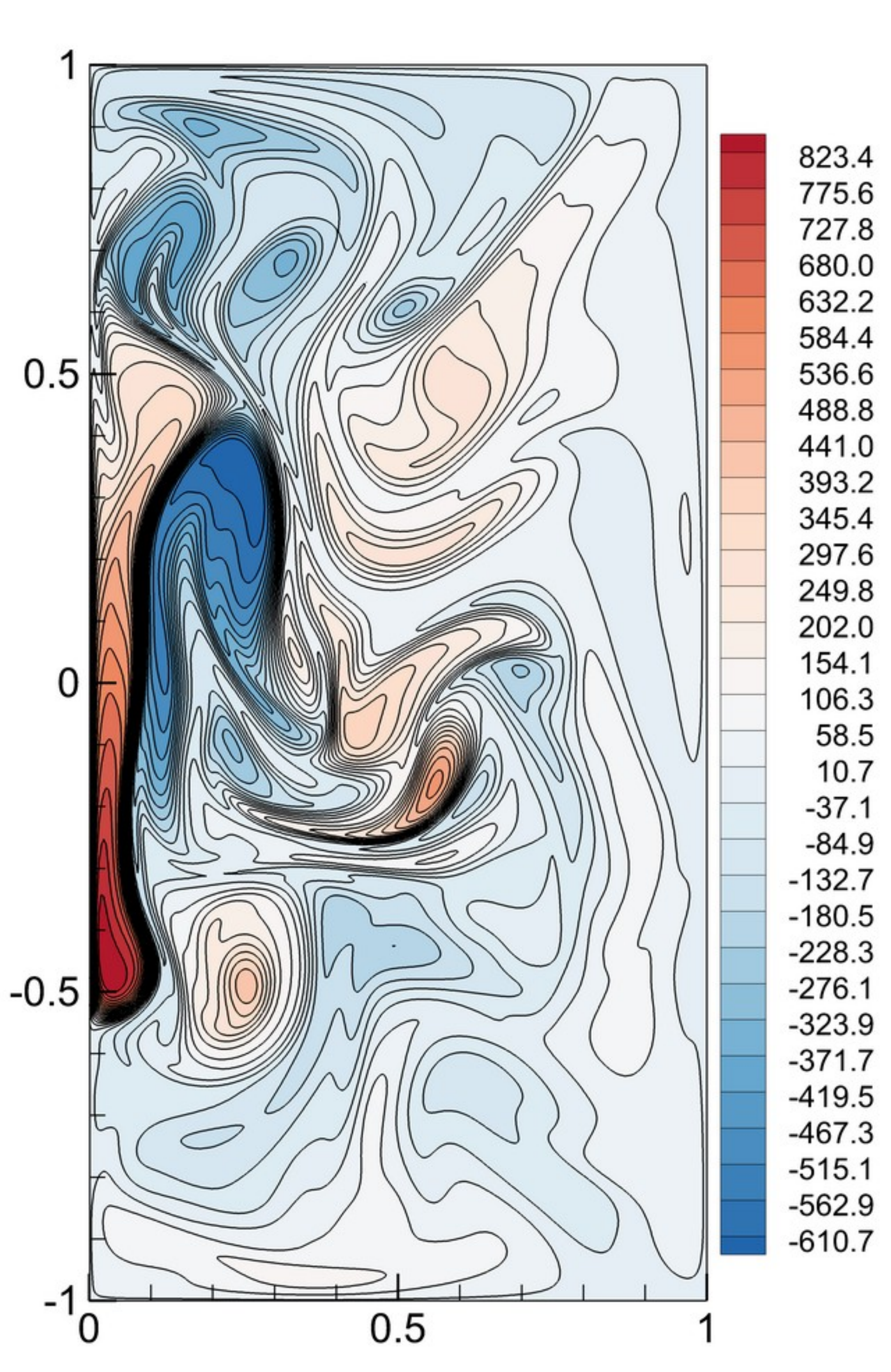}}
\subfigure[$\omega$ ($t=80$)]{\includegraphics[width=0.33\textwidth]{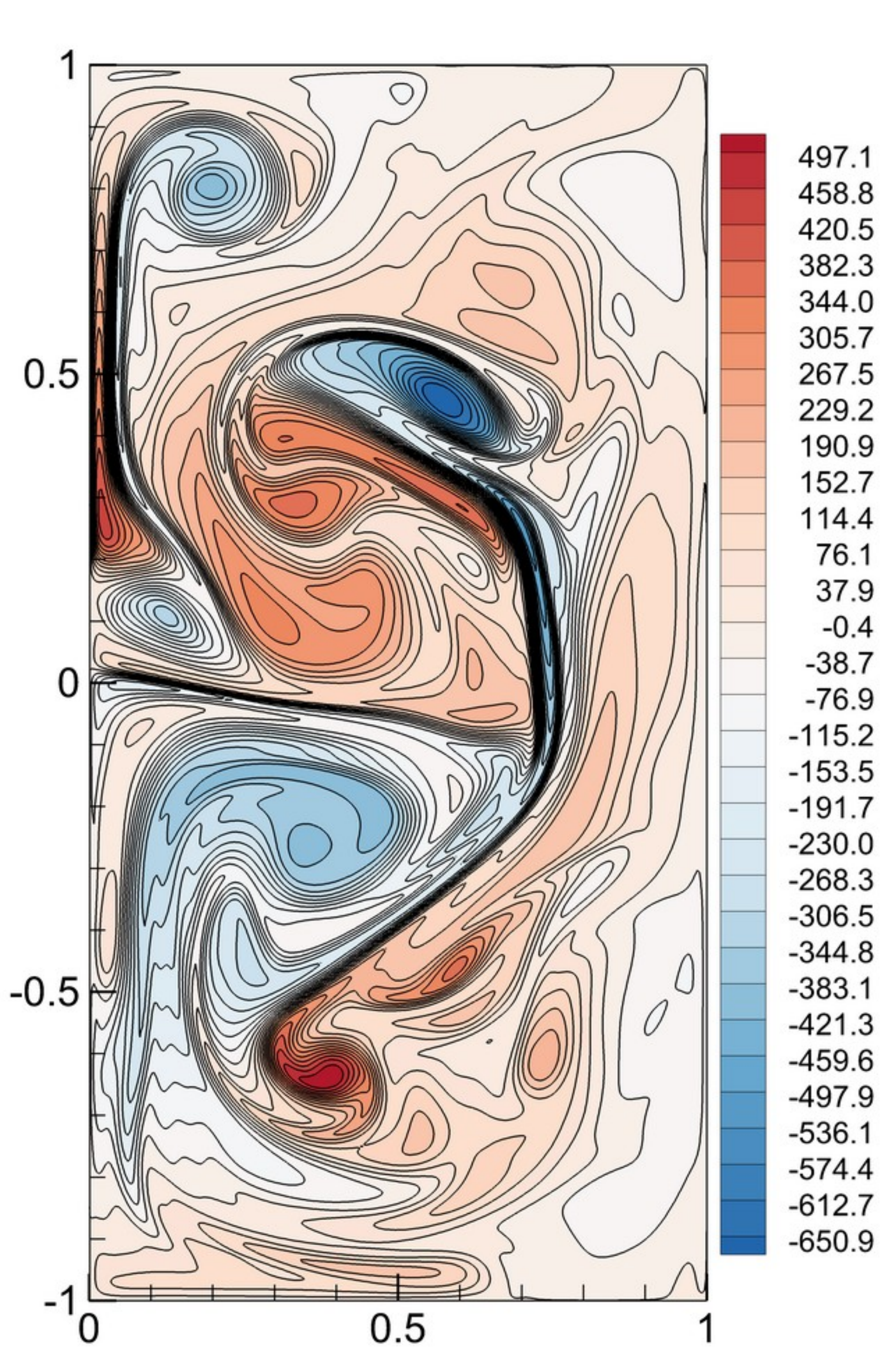}}
}\\
\mbox{
\subfigure[$\psi$ ($t=40$)]{\includegraphics[width=0.33\textwidth]{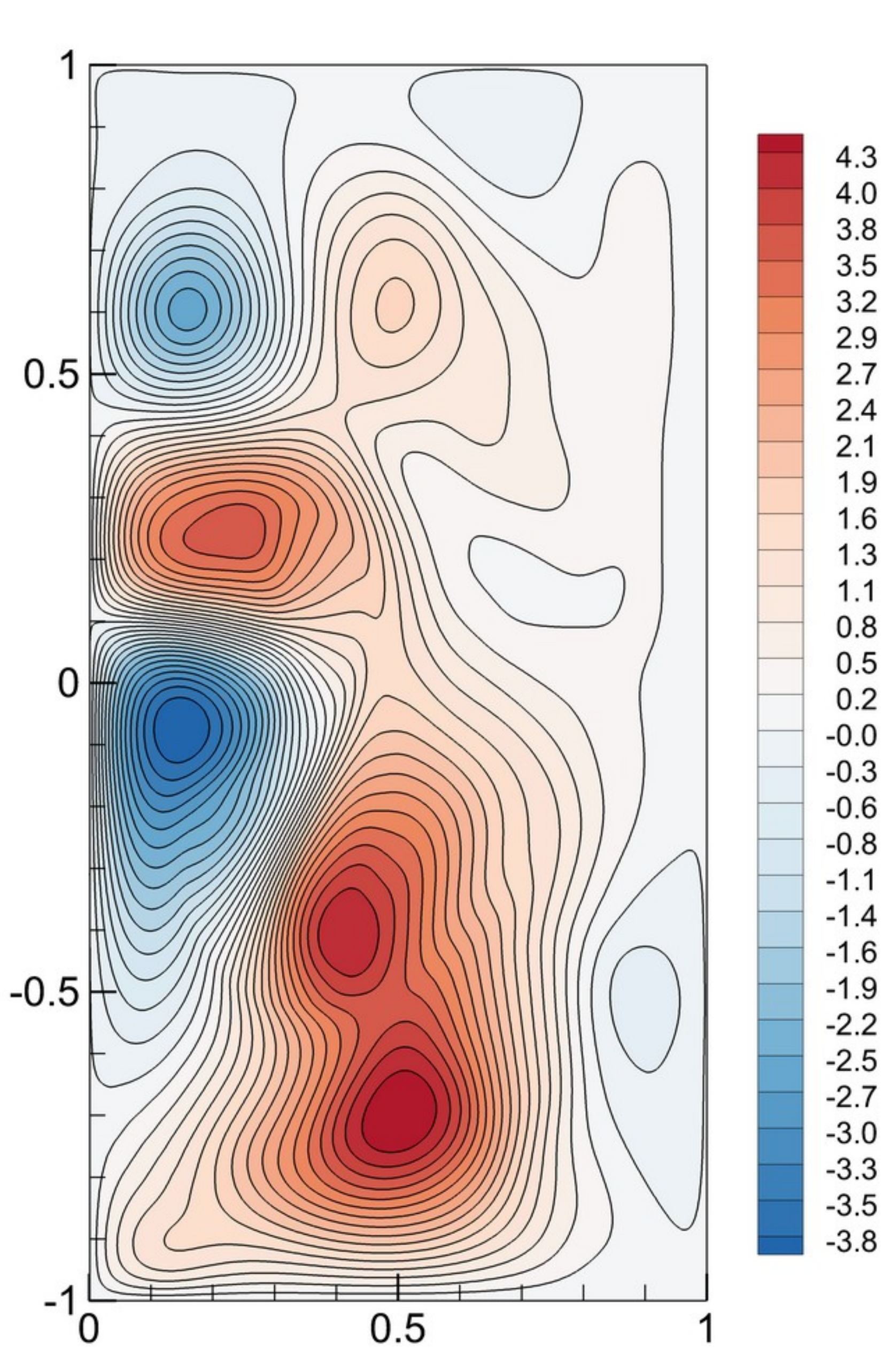}}
\subfigure[$\psi$ ($t=60$)]{\includegraphics[width=0.33\textwidth]{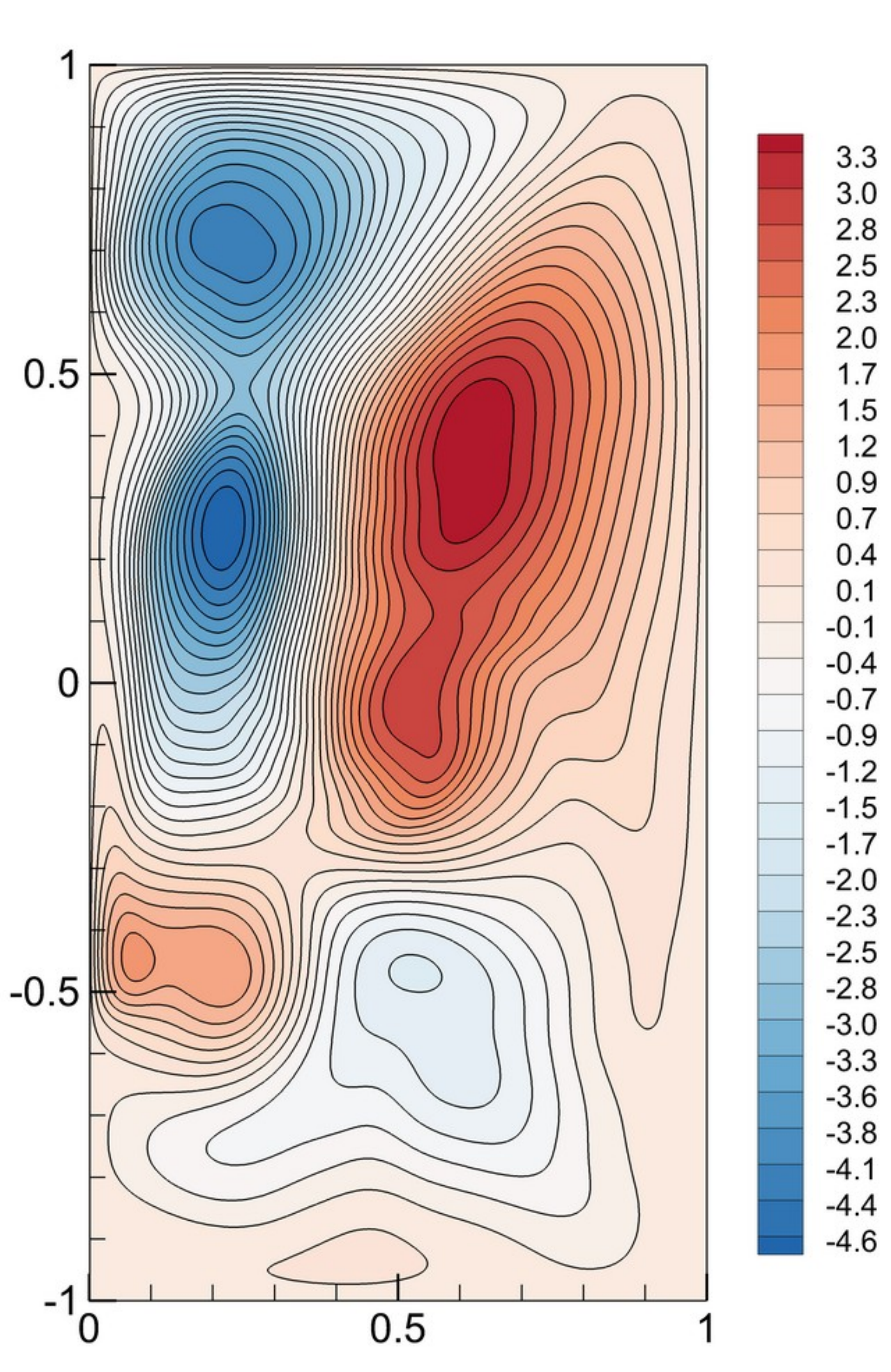}}
\subfigure[$\psi$ ($t=80$)]{\includegraphics[width=0.33\textwidth]{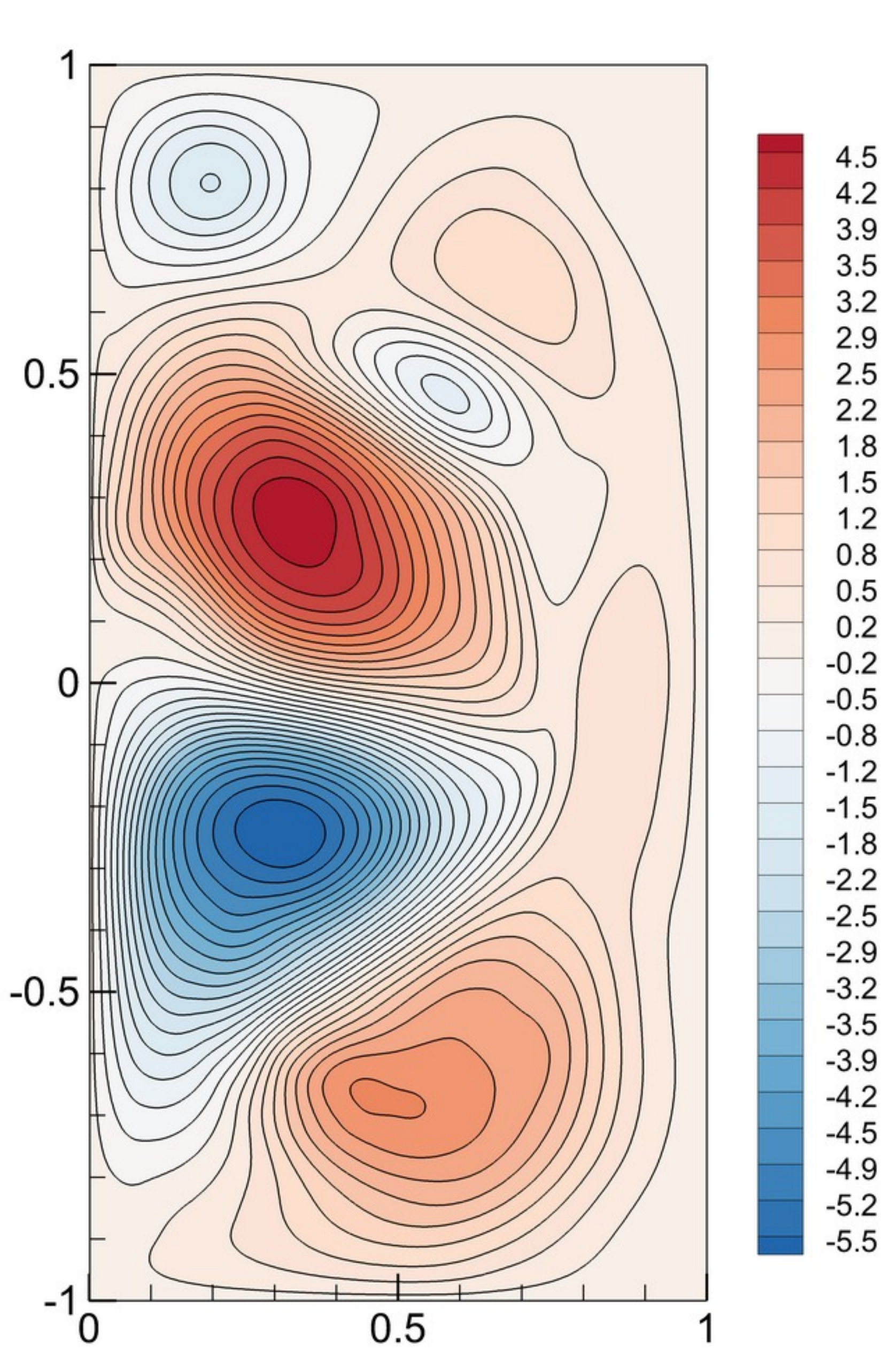}}
}
\caption{Instantaneous vorticity (a-c) and stream function (d-f) contour plots for Experiment 2.}
\label{fig:e2-ins}
\end{figure}

\begin{figure}
\includegraphics[width=1.0\textwidth]{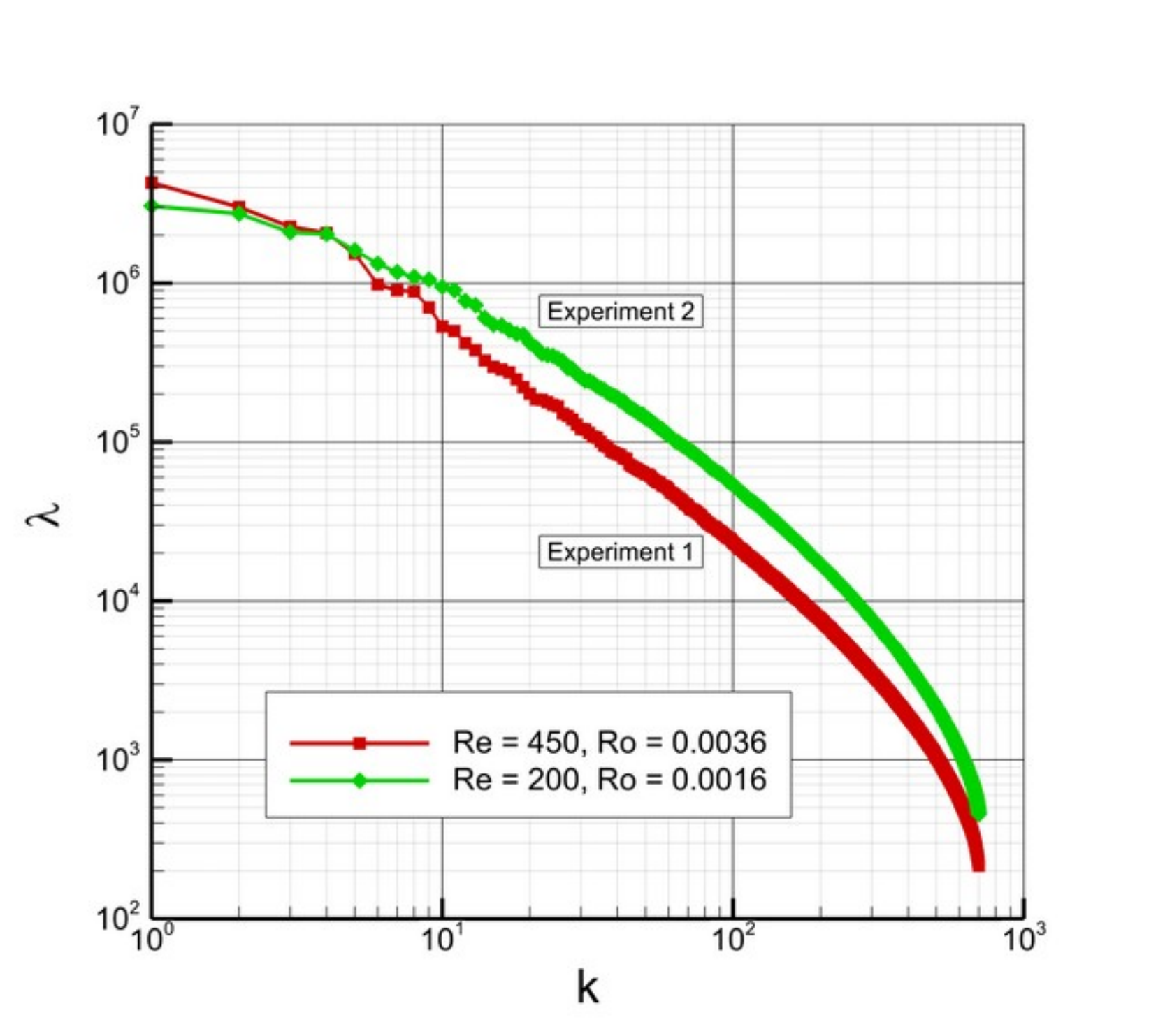}
\caption{Eigenvalues of the correlation matrix $C$ using 700 equally distributed snapshots between time $t=10$ and $t=80$.}
\label{fig:eig}
\end{figure}

\begin{figure}
\mbox{
\subfigure[$k=1$]{\includegraphics[width=0.33\textwidth]{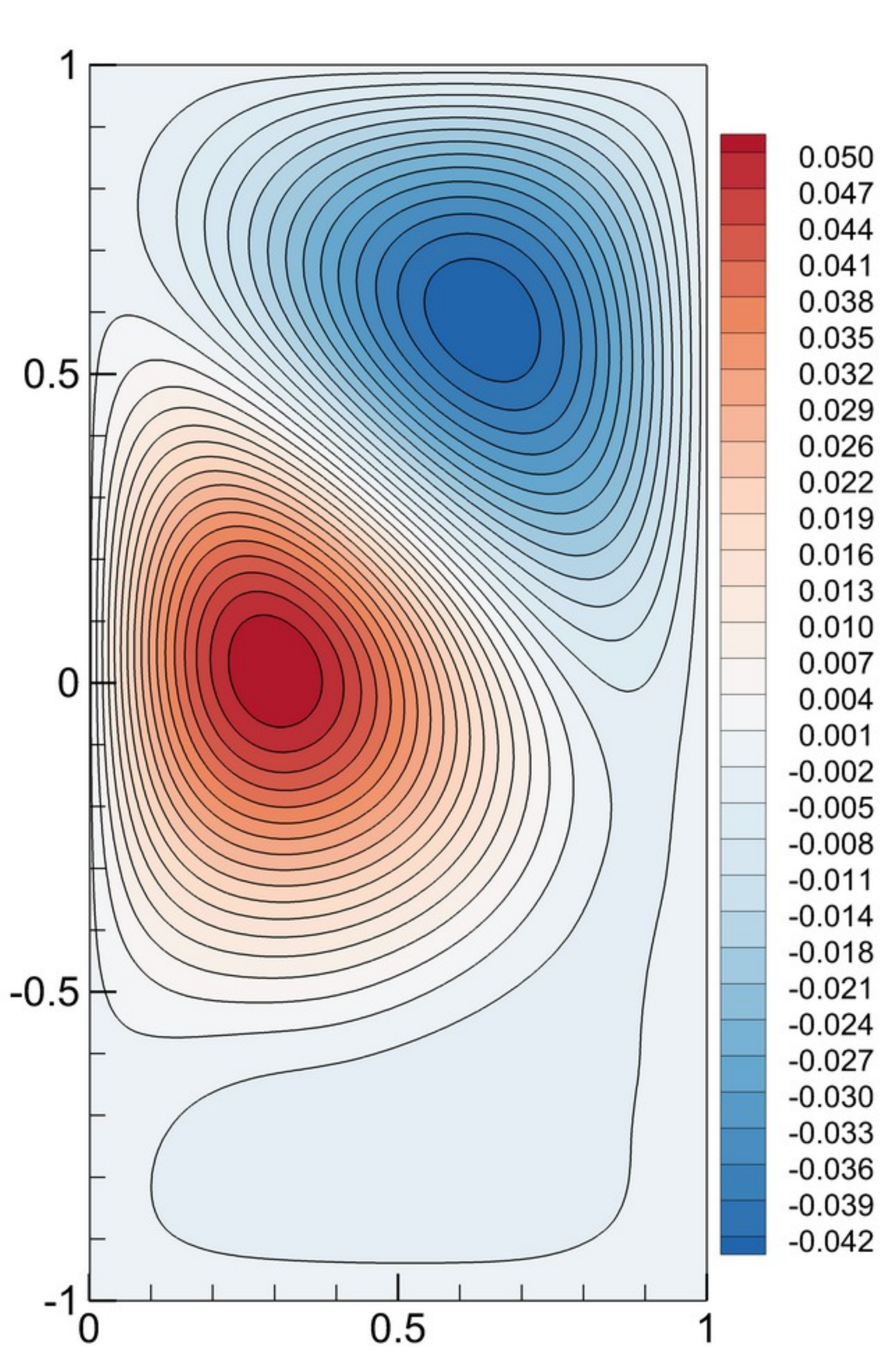}}
\subfigure[$k=5$]{\includegraphics[width=0.33\textwidth]{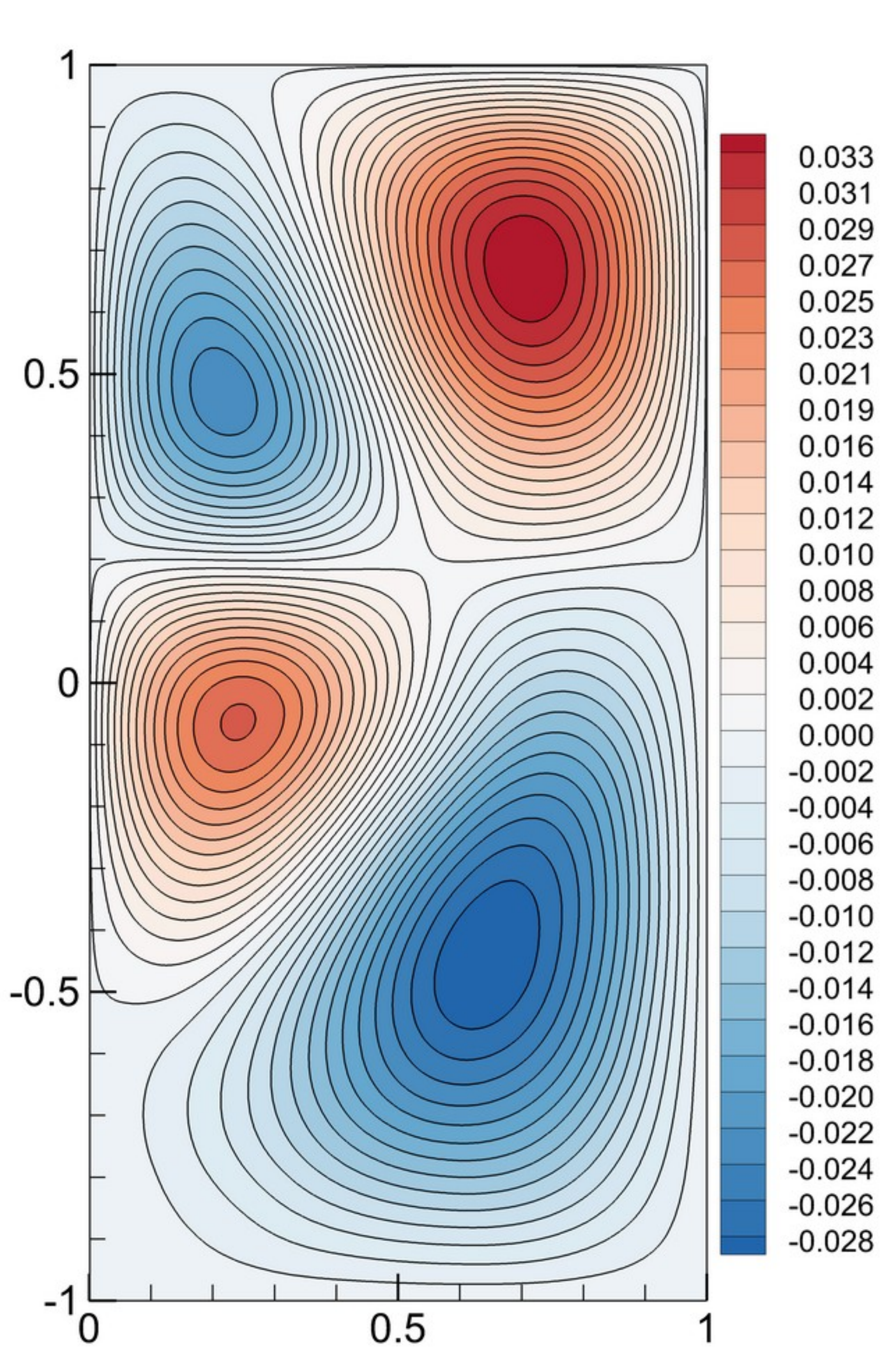}}
\subfigure[$k=10$]{\includegraphics[width=0.33\textwidth]{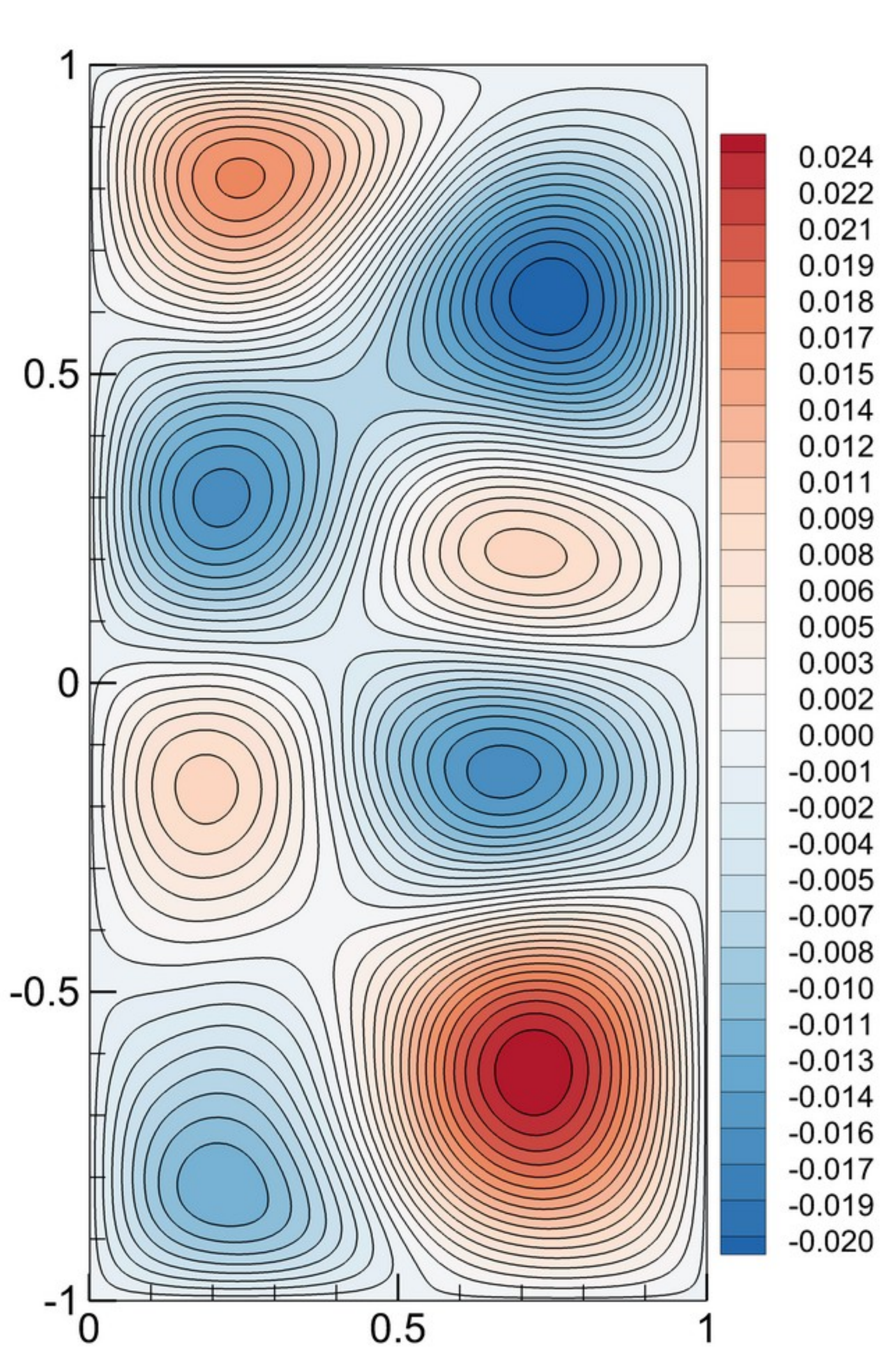}}
}\\
\mbox{
\subfigure[$k=15$]{\includegraphics[width=0.33\textwidth]{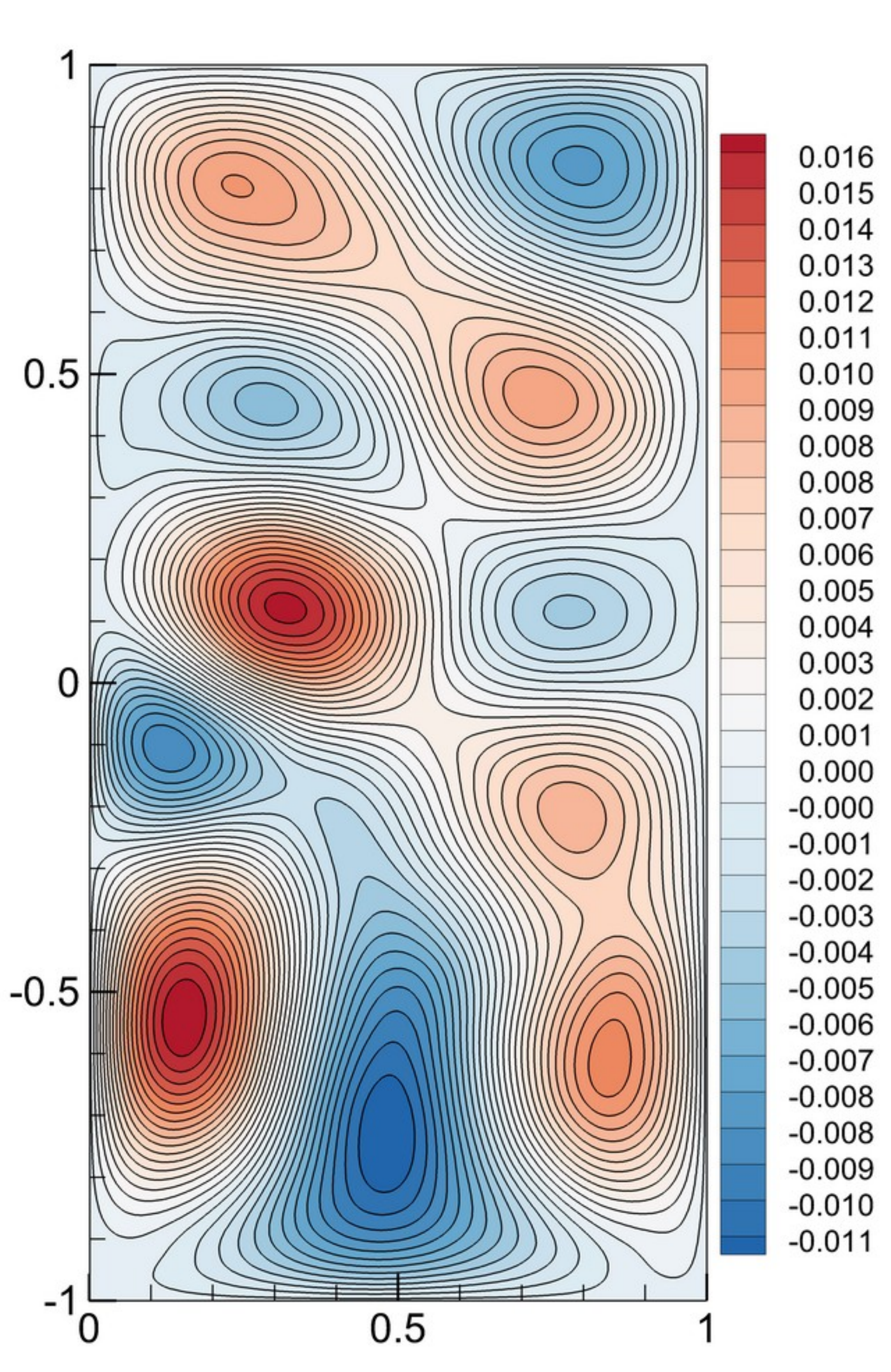}}
\subfigure[$k=20$]{\includegraphics[width=0.33\textwidth]{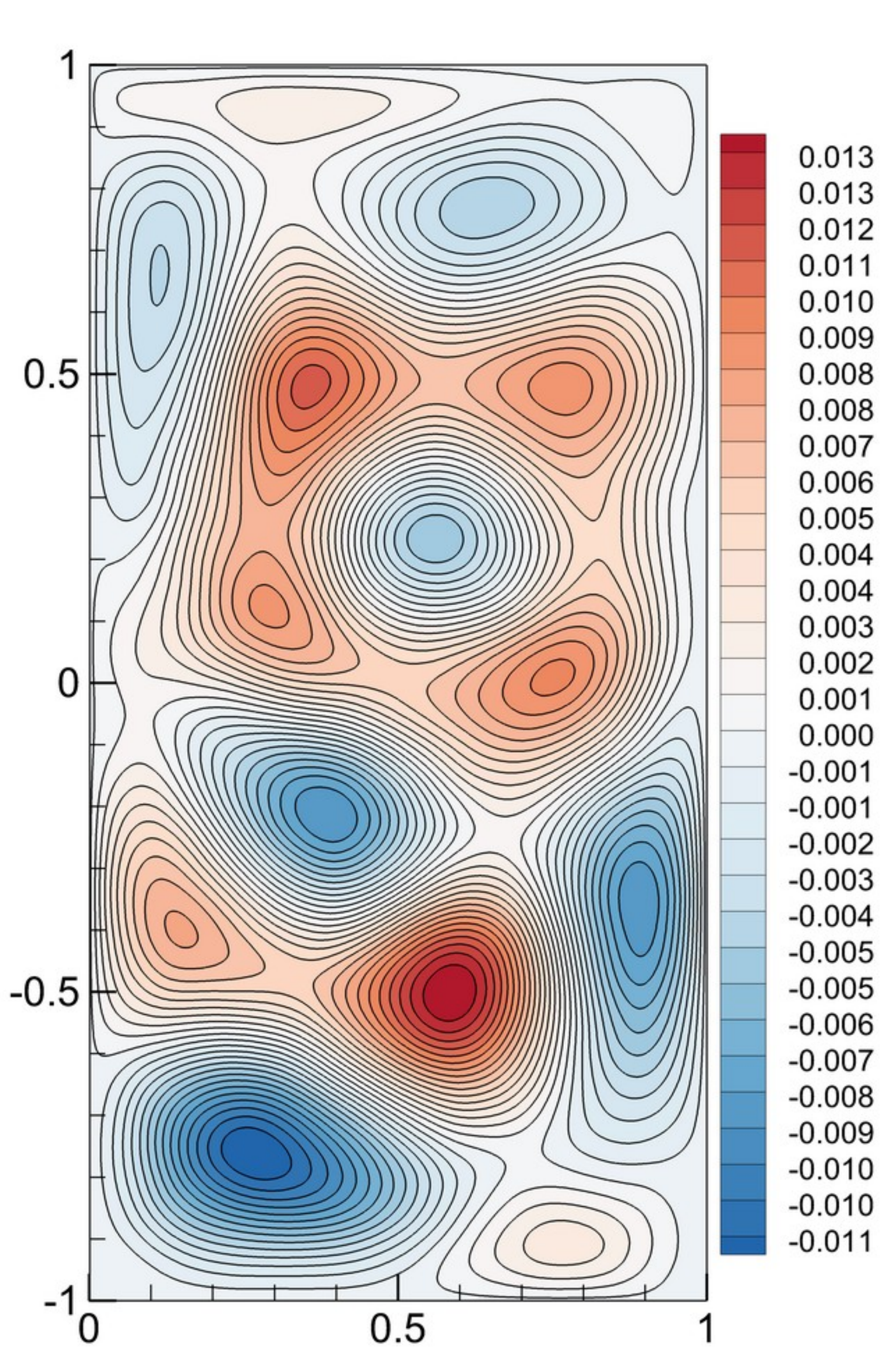}}
\subfigure[$k=30$]{\includegraphics[width=0.33\textwidth]{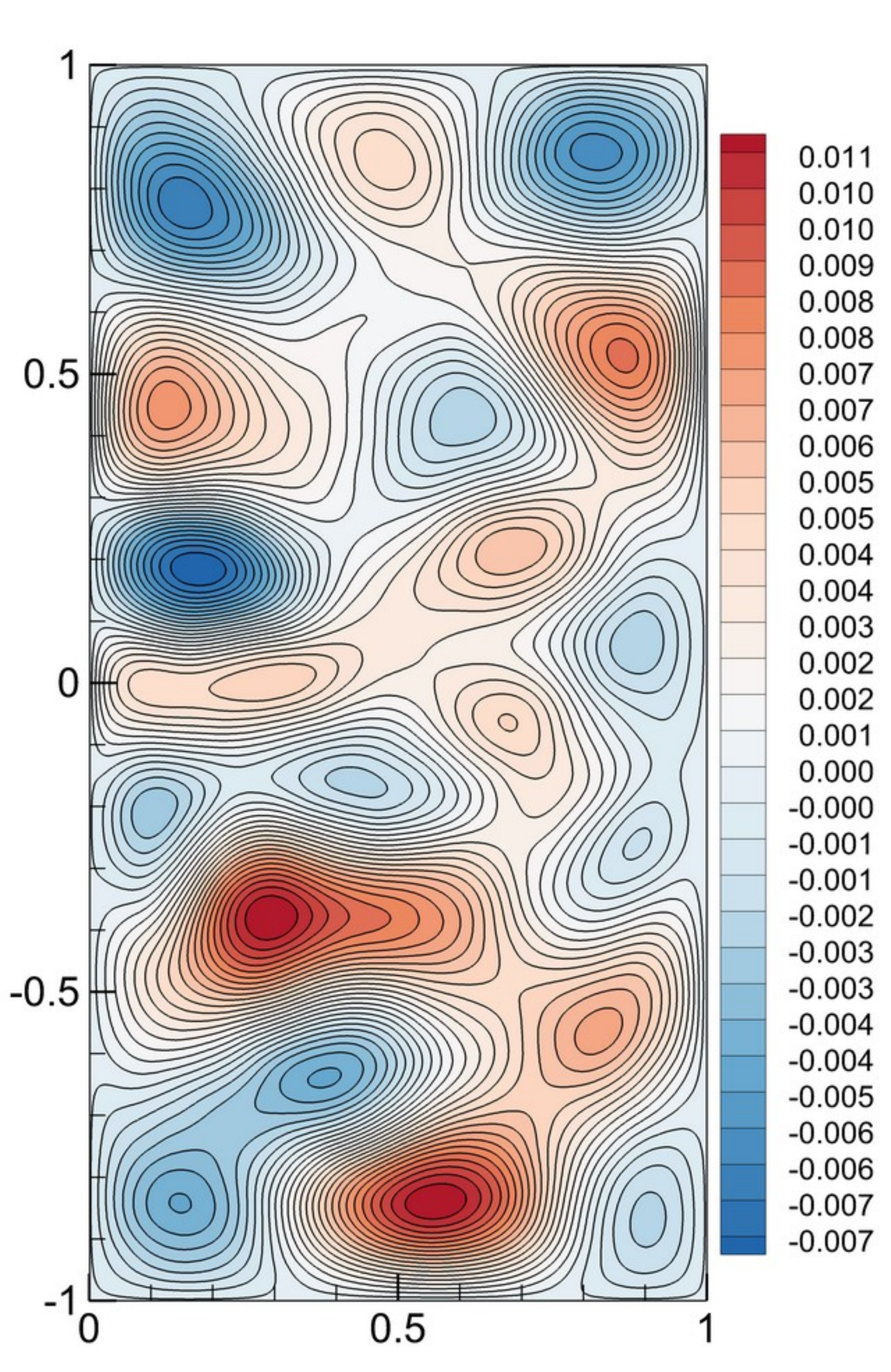}}
}
\caption{Illustrative examples of POD basis functions, $\varphi_{k}$, for Experiment 1.}
\label{fig:e1-bas}
\end{figure}

\begin{figure}
\mbox{
\subfigure[$k=1$]{\includegraphics[width=0.33\textwidth]{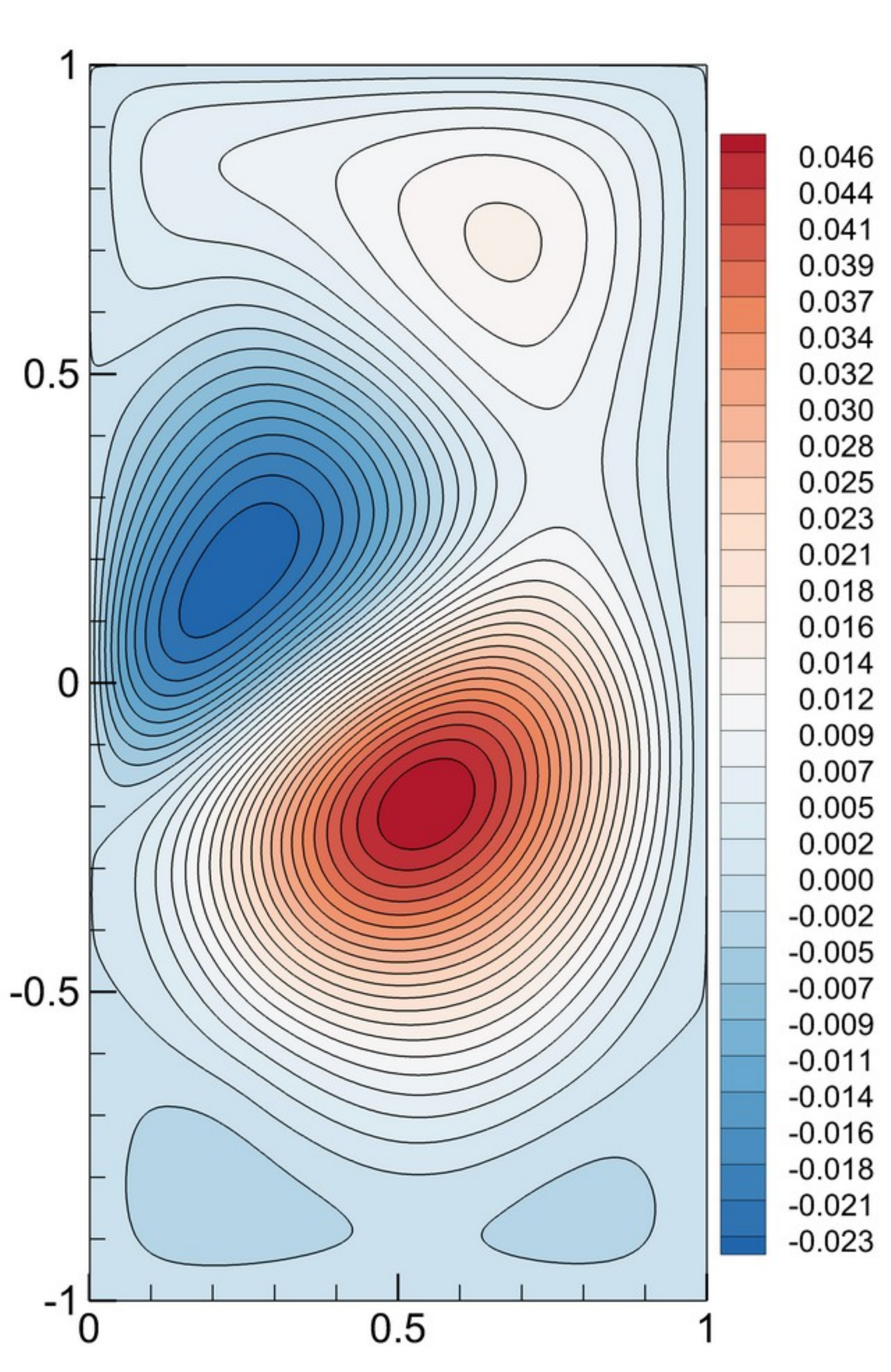}}
\subfigure[$k=5$]{\includegraphics[width=0.33\textwidth]{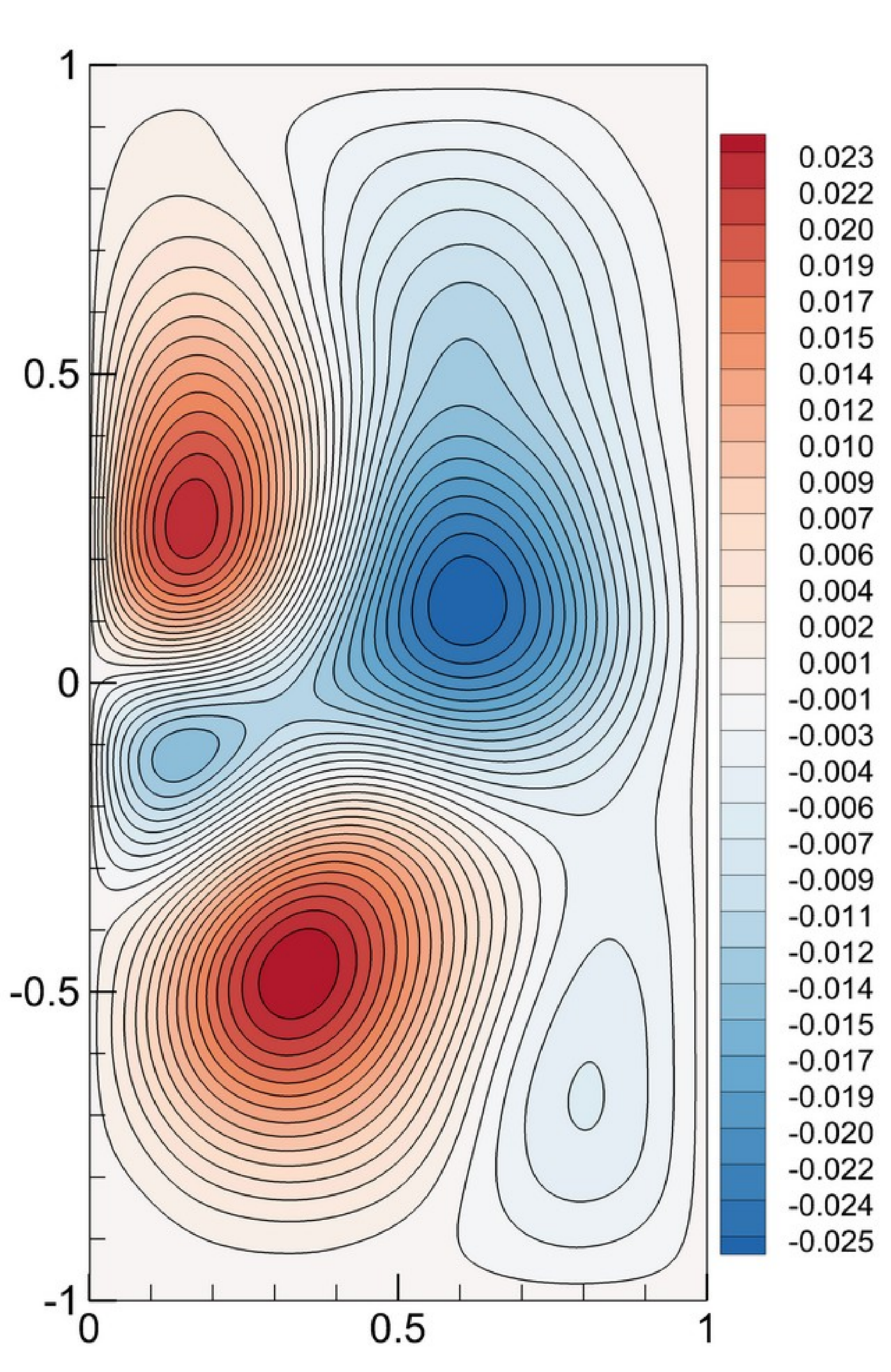}}
\subfigure[$k=10$]{\includegraphics[width=0.33\textwidth]{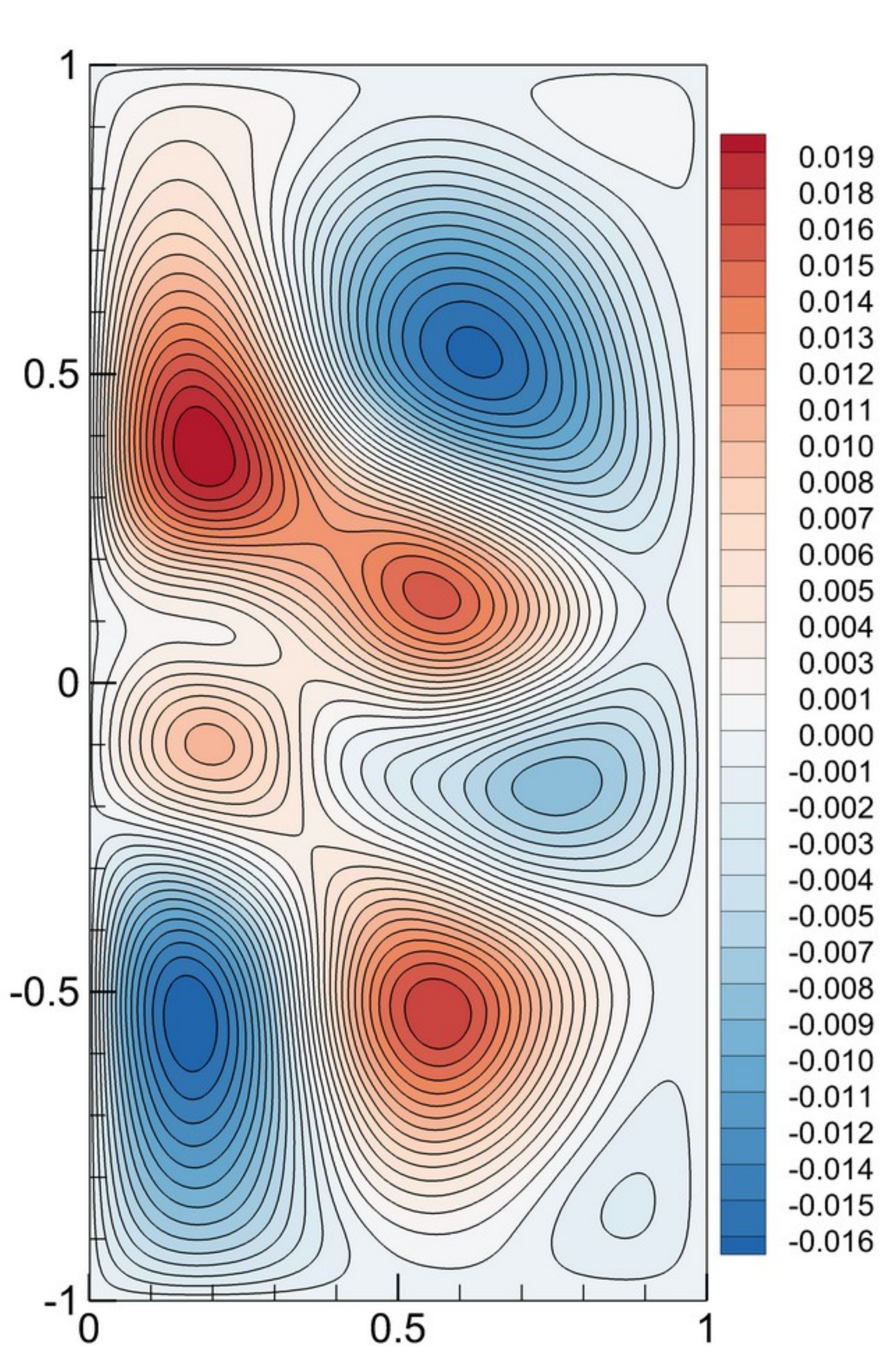}}
}\\
\mbox{
\subfigure[$k=15$]{\includegraphics[width=0.33\textwidth]{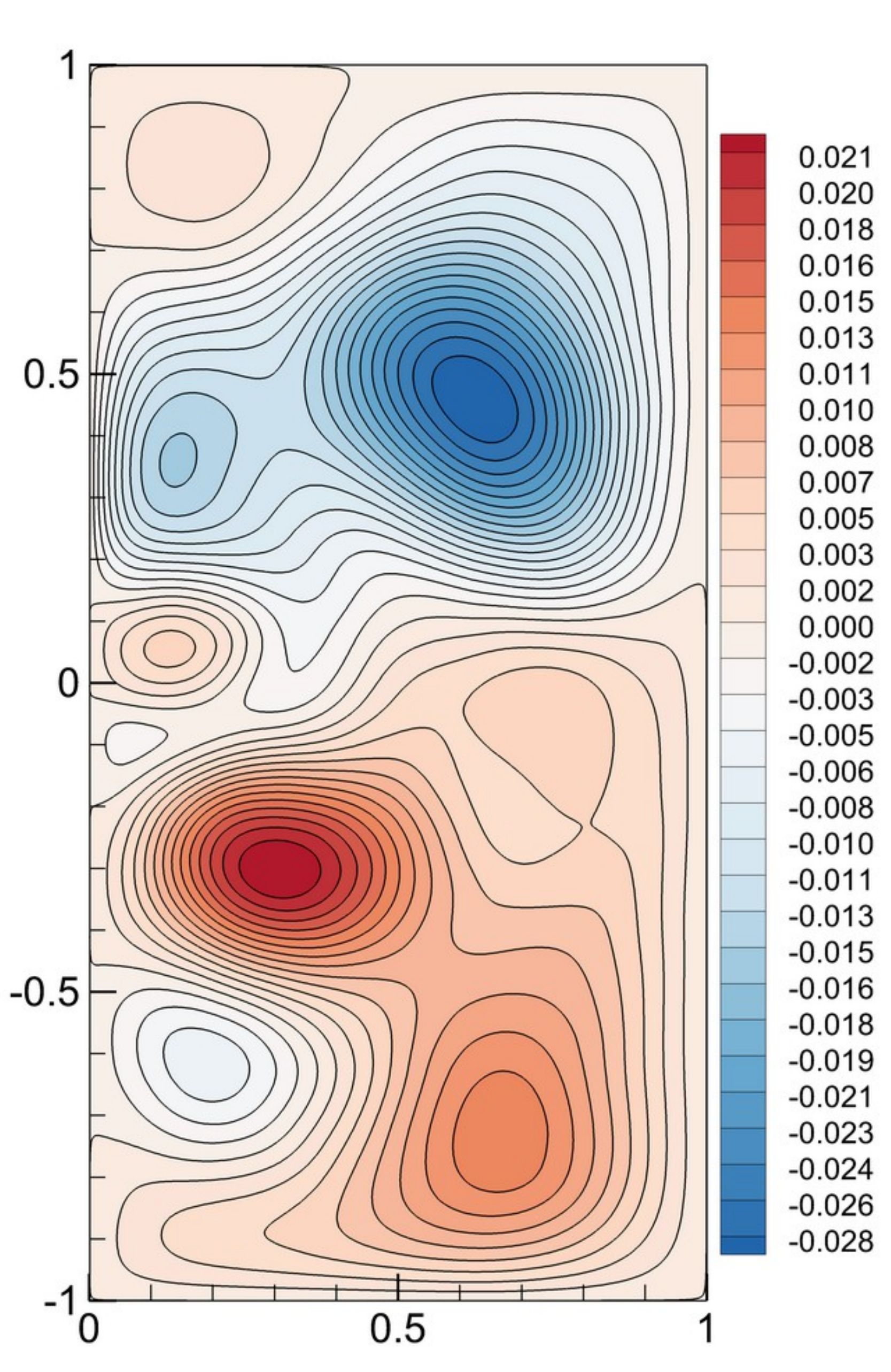}}
\subfigure[$k=20$]{\includegraphics[width=0.33\textwidth]{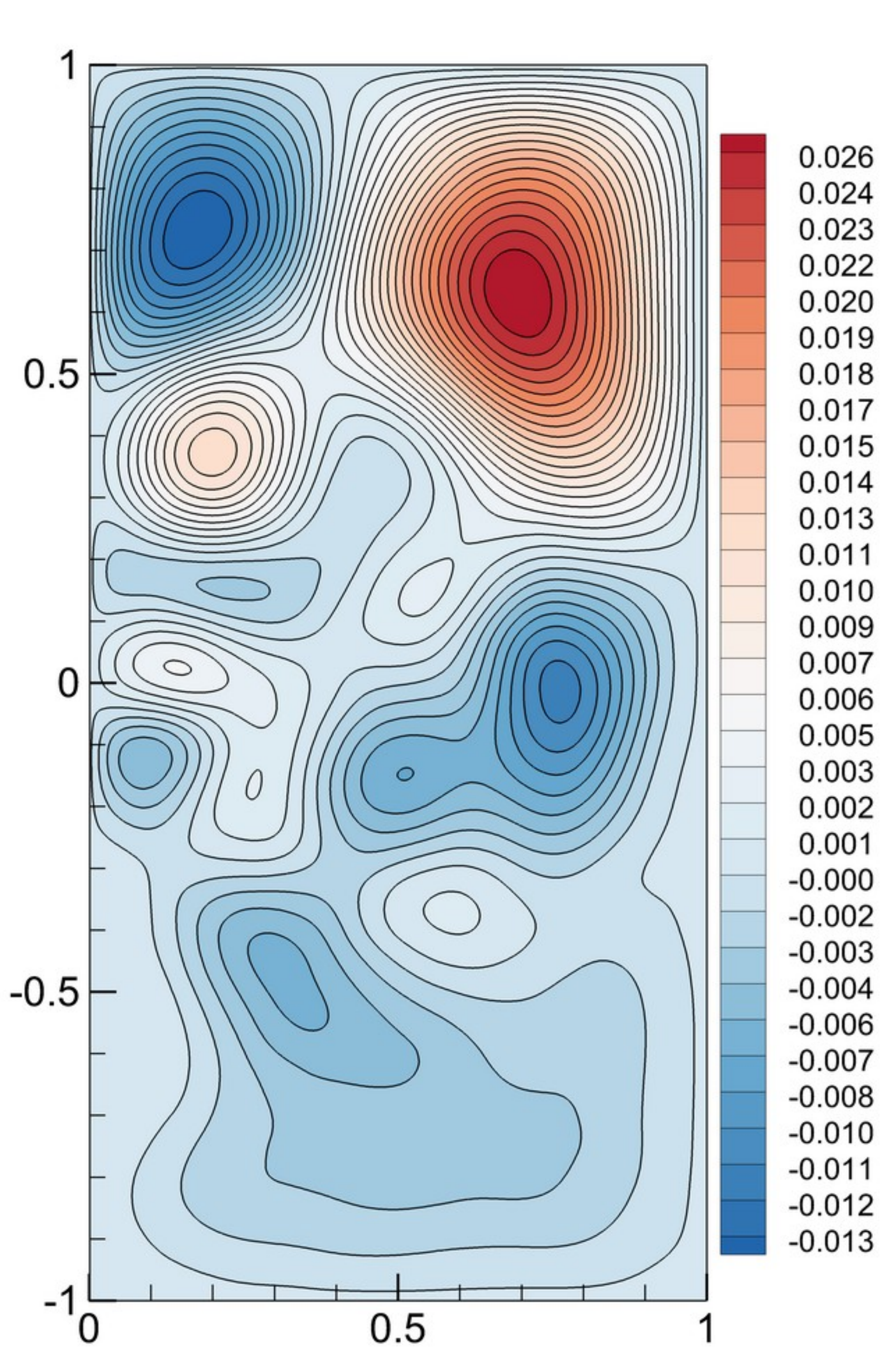}}
\subfigure[$k=30$]{\includegraphics[width=0.33\textwidth]{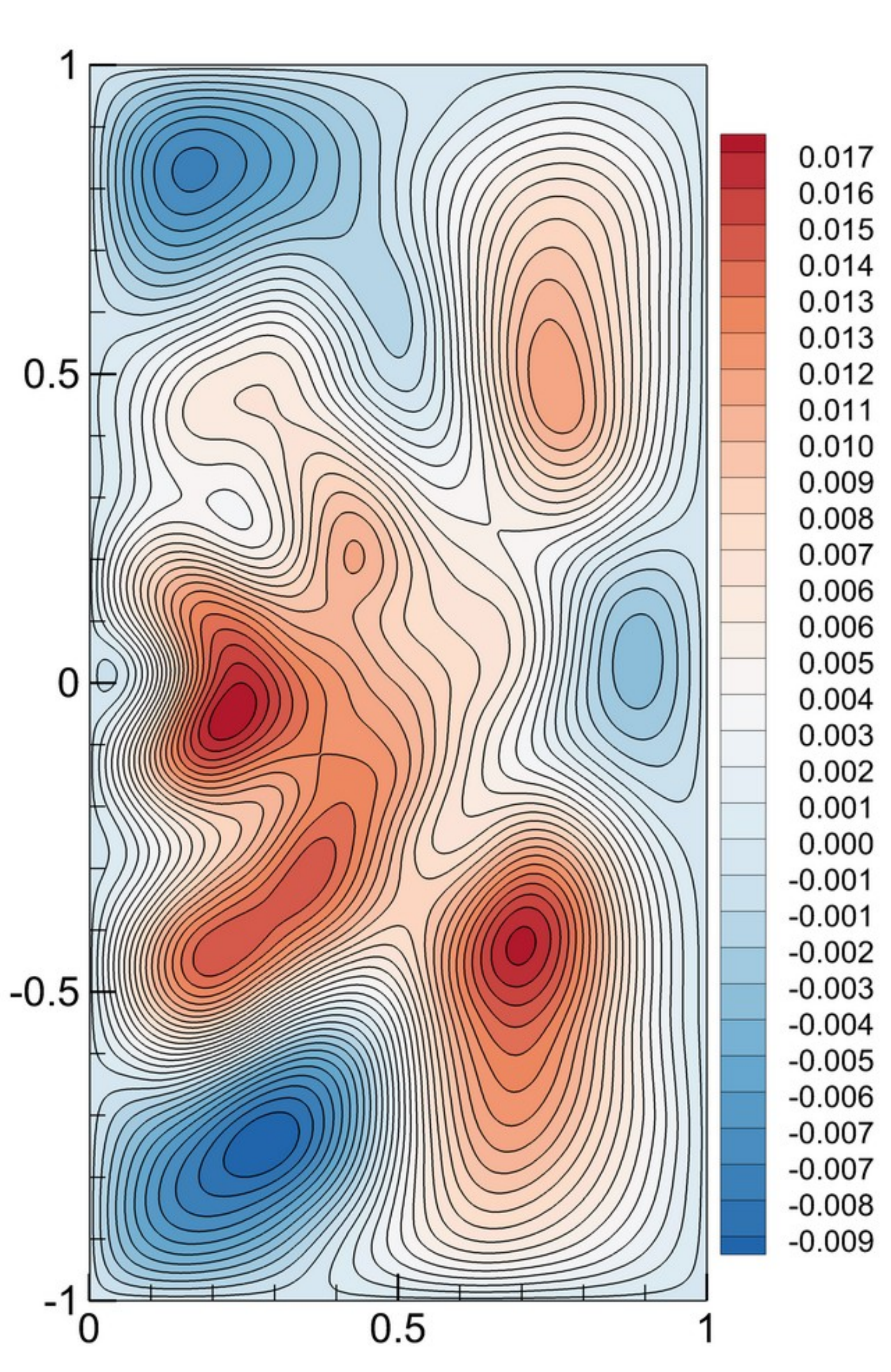}}
}
\caption{Illustrative examples of POD basis functions, $\varphi_{k}$, for Experiment 2.}
\label{fig:e2-bas}
\end{figure}

\begin{figure}
\includegraphics[width=1.0\textwidth]{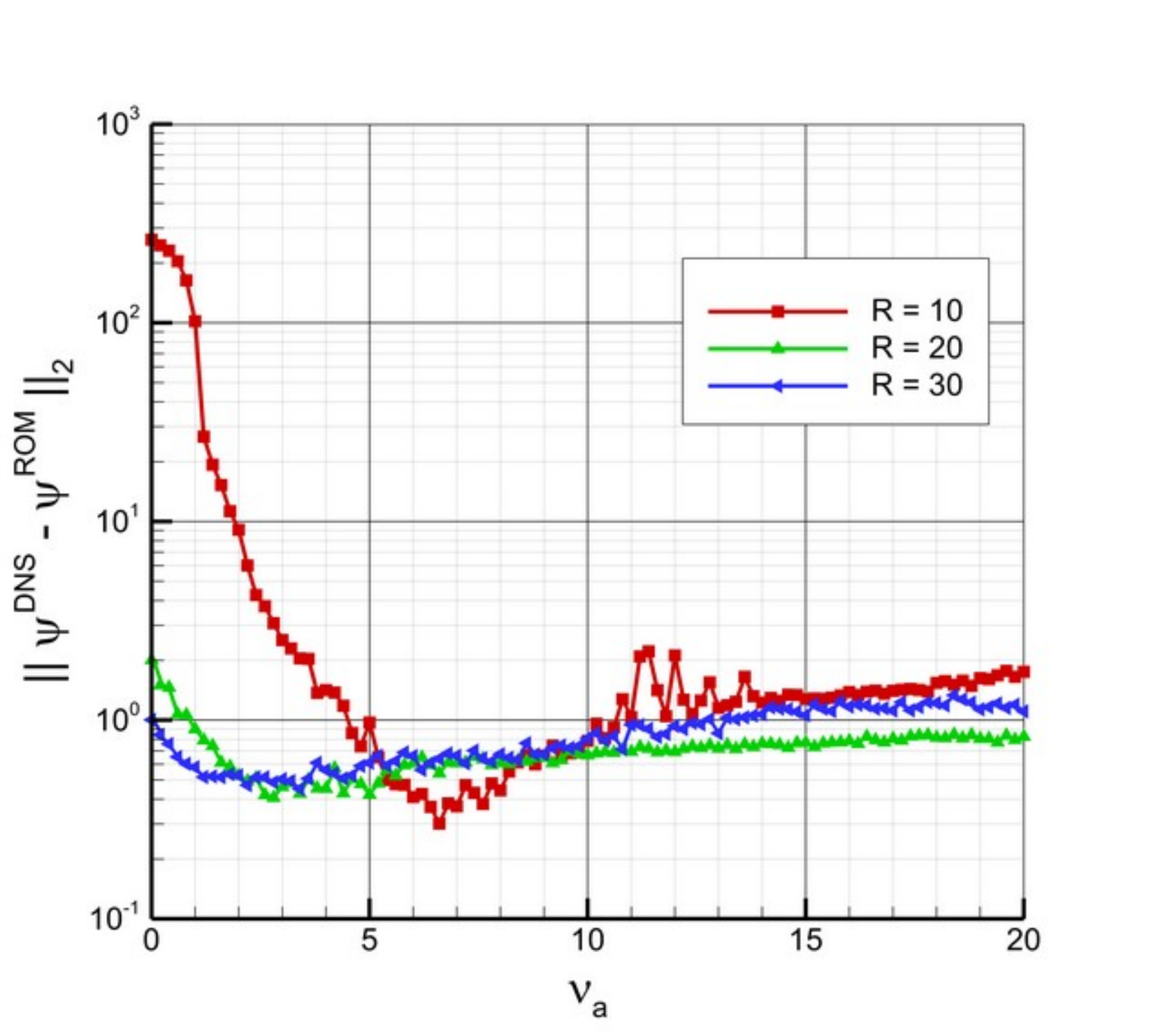}
\caption{Sensitivity analysis for Experiment 1 showing the mean stream function $L^2$-norm with respect to the eddy viscosity stabilization parameter given in Eq.~(\ref{eq:rem}). Error norms are computed using the DNS reference solution for three different POD-ROMs with different numbers of modes $R$.  Note that $\nu_a=0$ corresponds to the standard Galerkin POD-ROM.}
\label{fig:e1-sens}
\end{figure}

\begin{figure}
\includegraphics[width=1.0\textwidth]{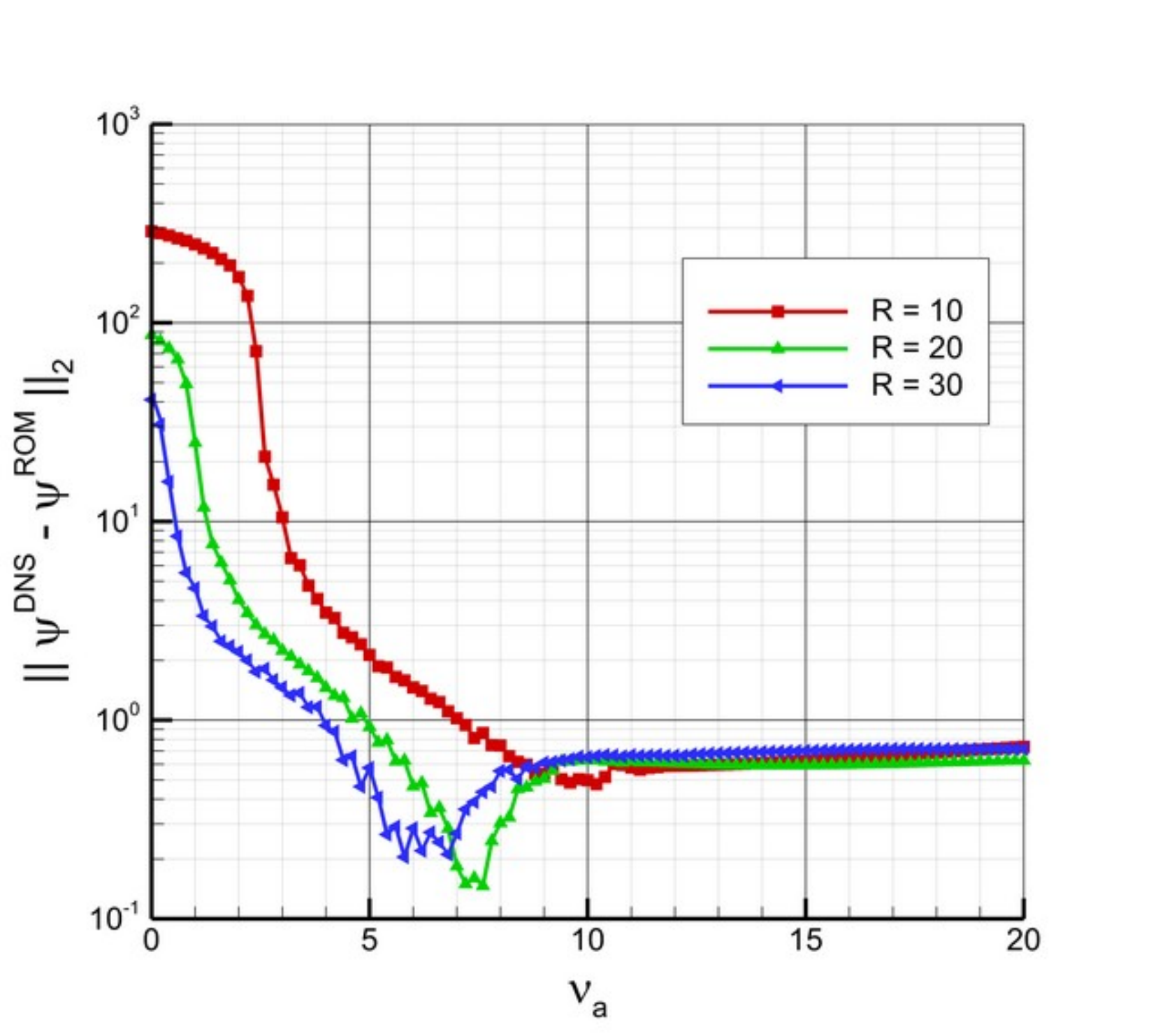}
\caption{Sensitivity analysis for Experiment 2 showing the mean stream function $L^2$-norm with respect to the eddy viscosity stabilization parameter given in Eq.~(\ref{eq:rem}). Error norms are computed using the DNS reference solution for three different reduced-order models with different number of modes $R$. Note that $\nu_a=0$ correspond standard Galerkin POD-ROM.}
\label{fig:e2-sens}
\end{figure}

\begin{figure}
\mbox{
\subfigure[DNS ($256 \times 512$)]{\includegraphics[width=0.33\textwidth]{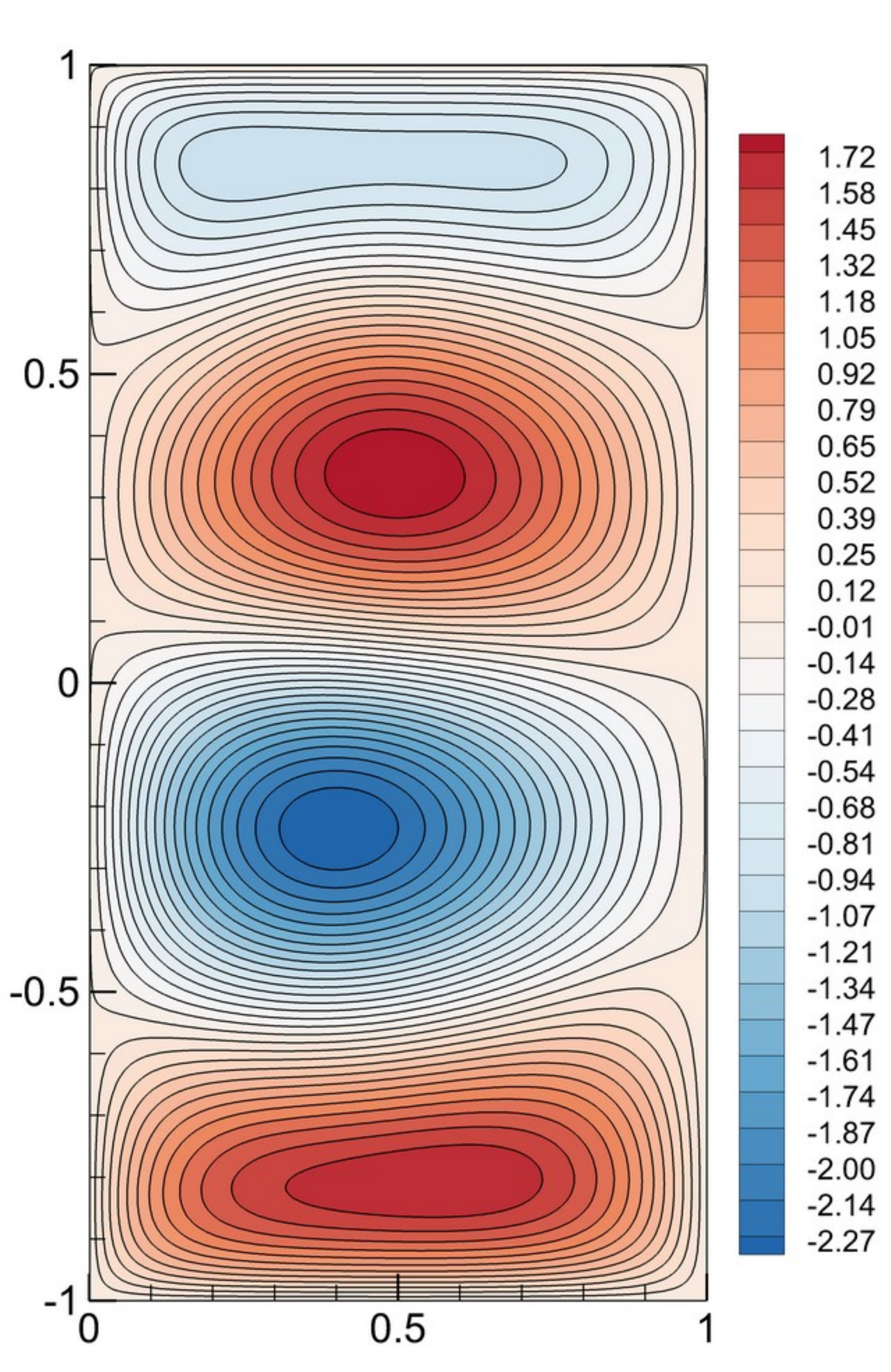}}
\subfigure[Galerkin POD-ROM ($R=10$)]{\includegraphics[width=0.33\textwidth]{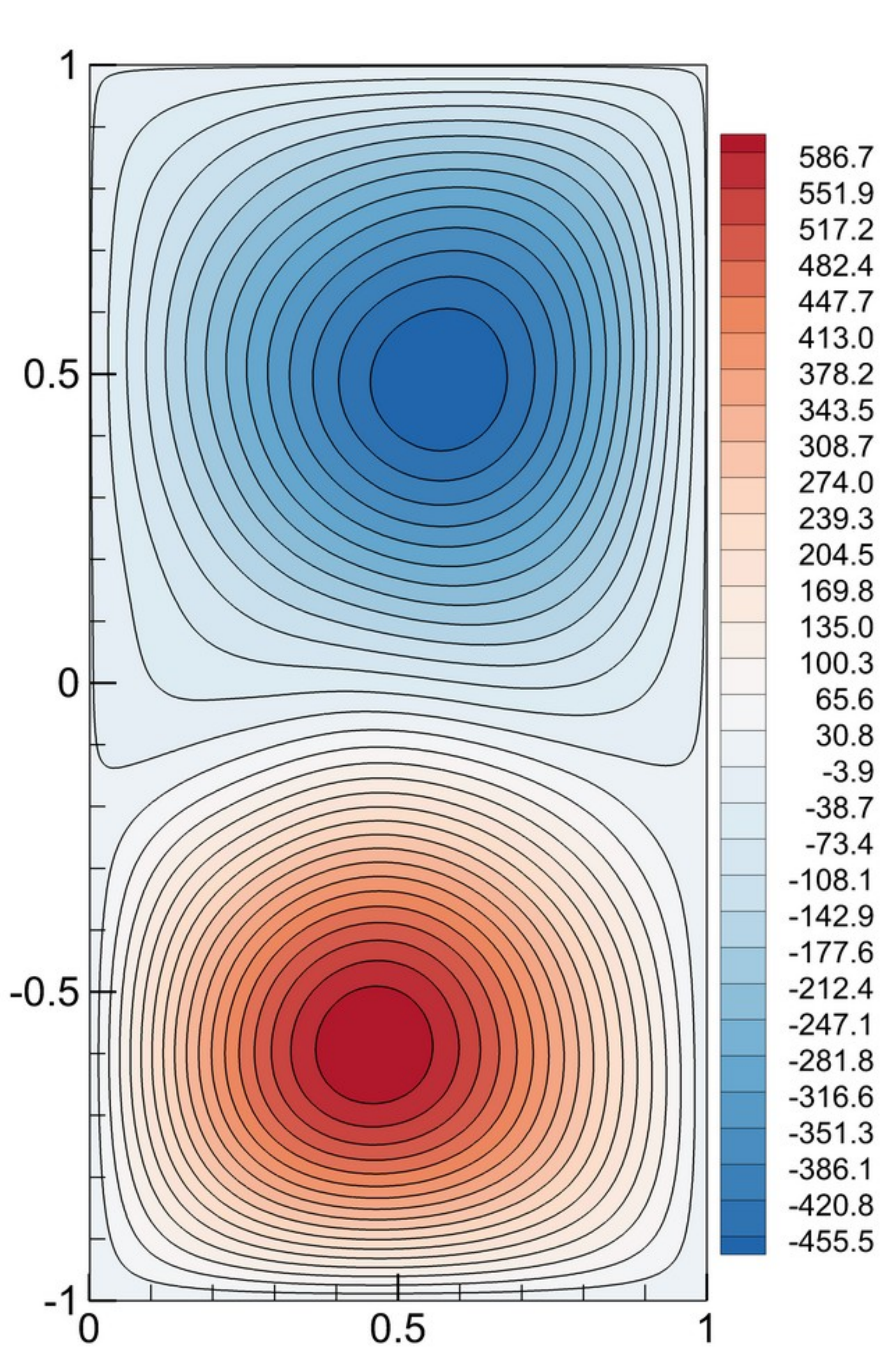}}
\subfigure[Stabilized POD-ROM ($R=10$)]{\includegraphics[width=0.33\textwidth]{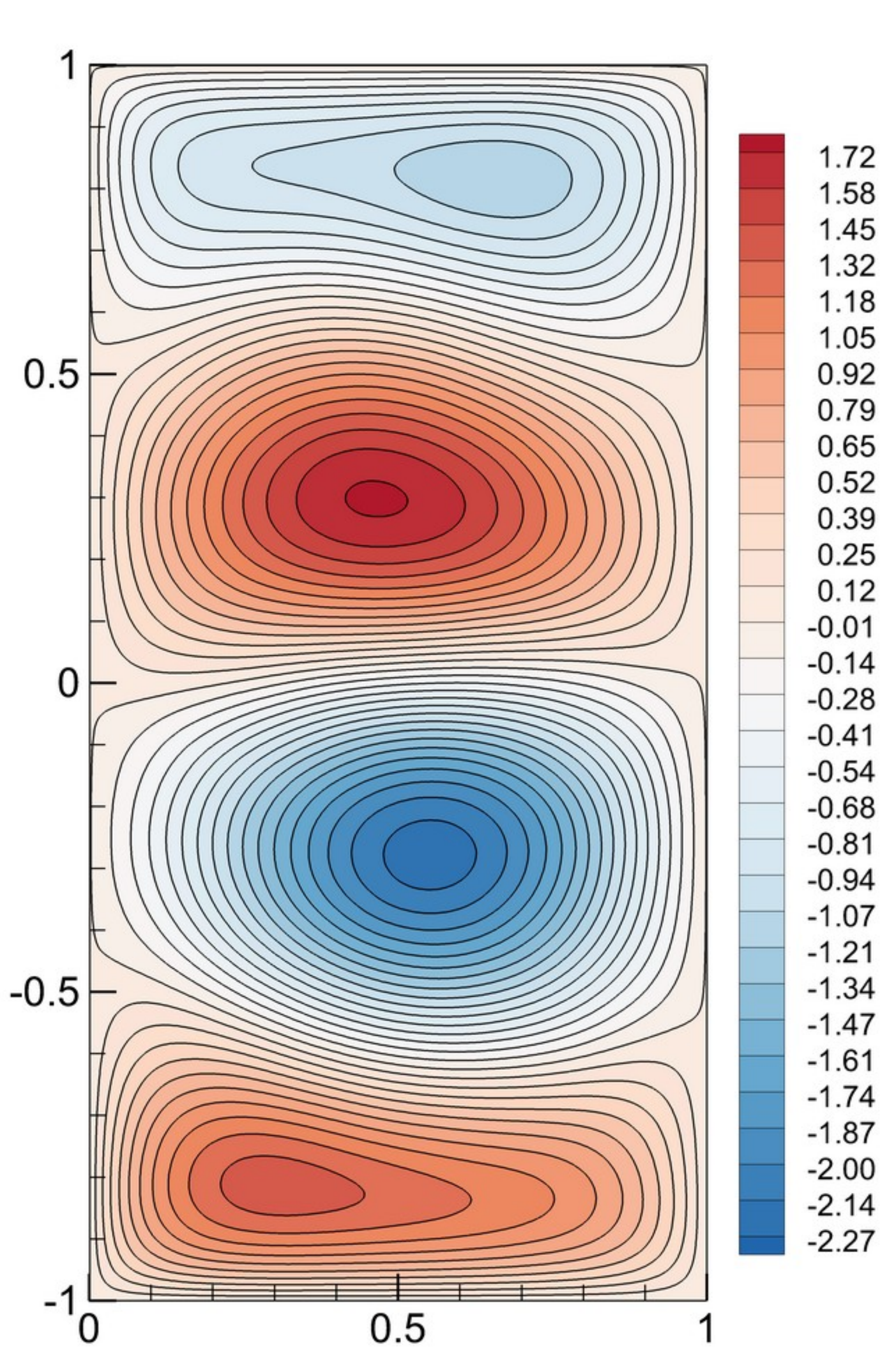}}
}
\caption{Comparison of the mean stream function contour plots for Experiment 1: (a) reference DNS computation at a resolution of $256 \times 512$ (with a computational cost of 326 hours of running CPU time); (b) standard Galerkin POD-ROM without any stabilization using $R=10$ modes (with a computational cost of 85 seconds of running CPU time); and (c) stabilized POD-ROM with $\nu_a=6.6$ using $R=10$ modes (with a computational cost of 85 seconds of running CPU time). Note that the standard Galerkin POD-ROM yields nonphysical result, whereas the DNS and the stabilized POD-ROM model results are qualitatively close. The contour interval layouts are identical only for (a) and (c).}
\label{fig:e1-com}
\end{figure}

\begin{figure}
\mbox{
\subfigure[DNS ($256 \times 512$)]{\includegraphics[width=0.33\textwidth]{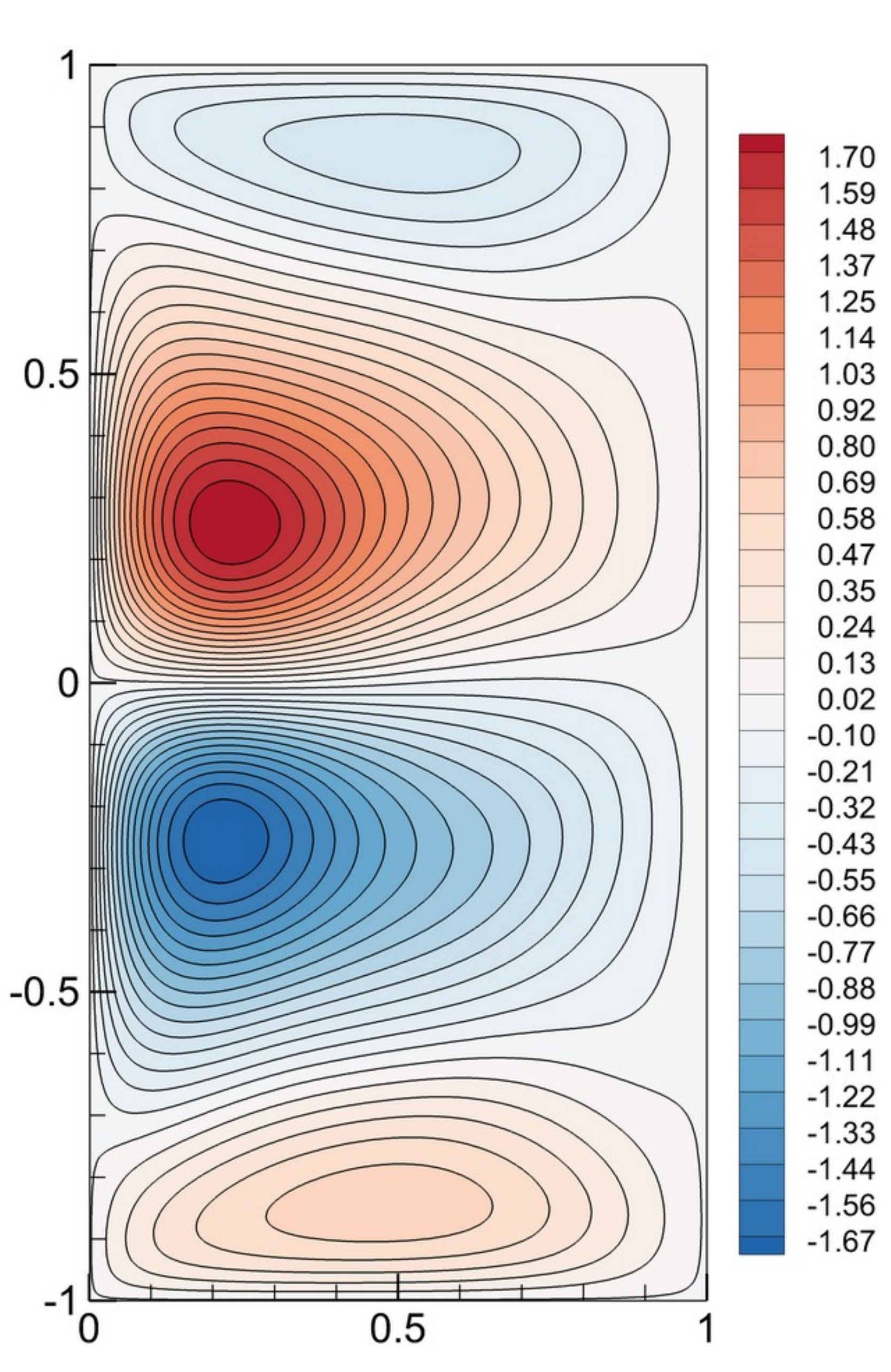}}
\subfigure[Galerkin POD-ROM ($R=20$)]{\includegraphics[width=0.33\textwidth]{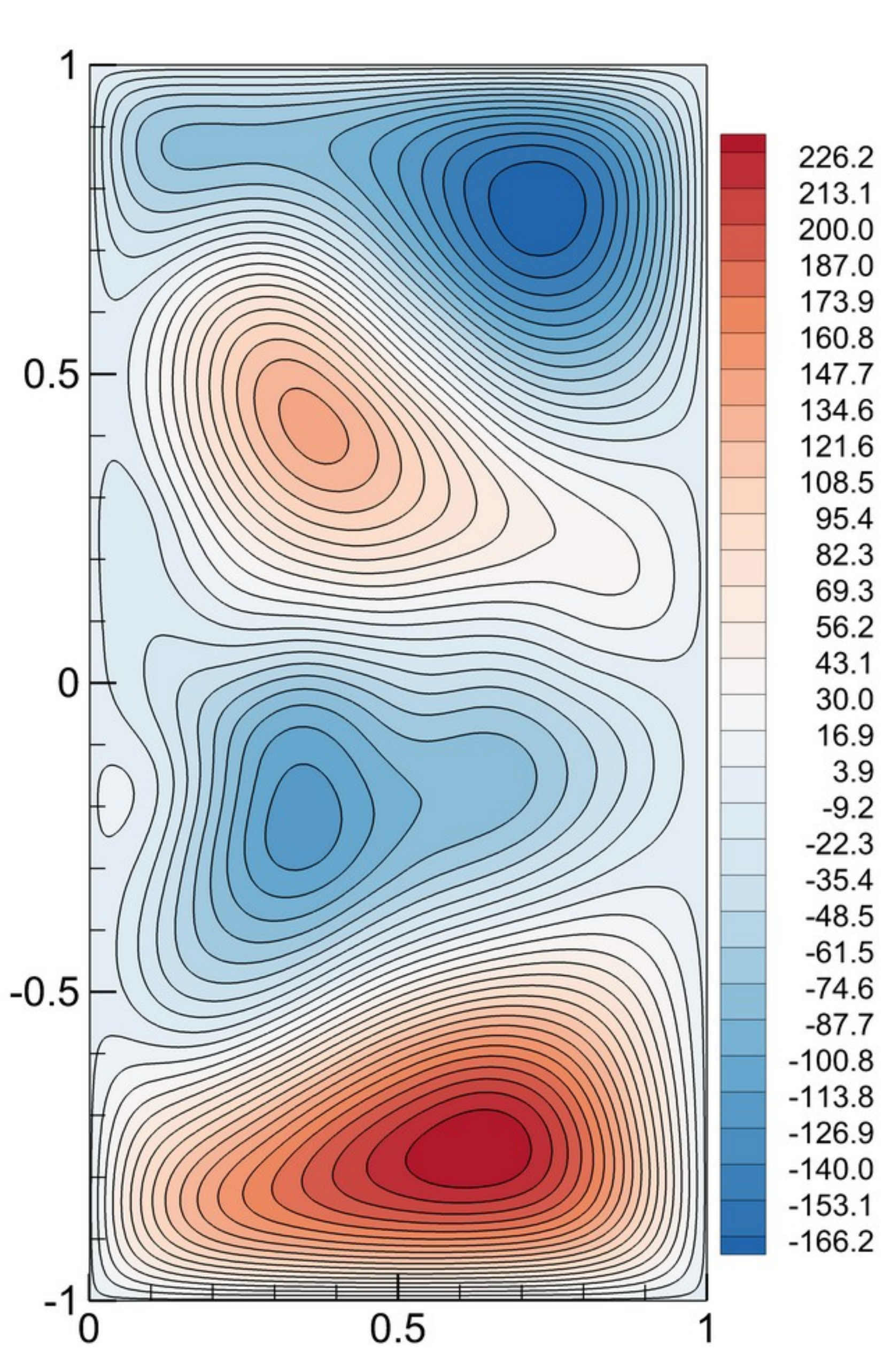}}
\subfigure[Stabilized POD-ROM ($R=20$)]{\includegraphics[width=0.33\textwidth]{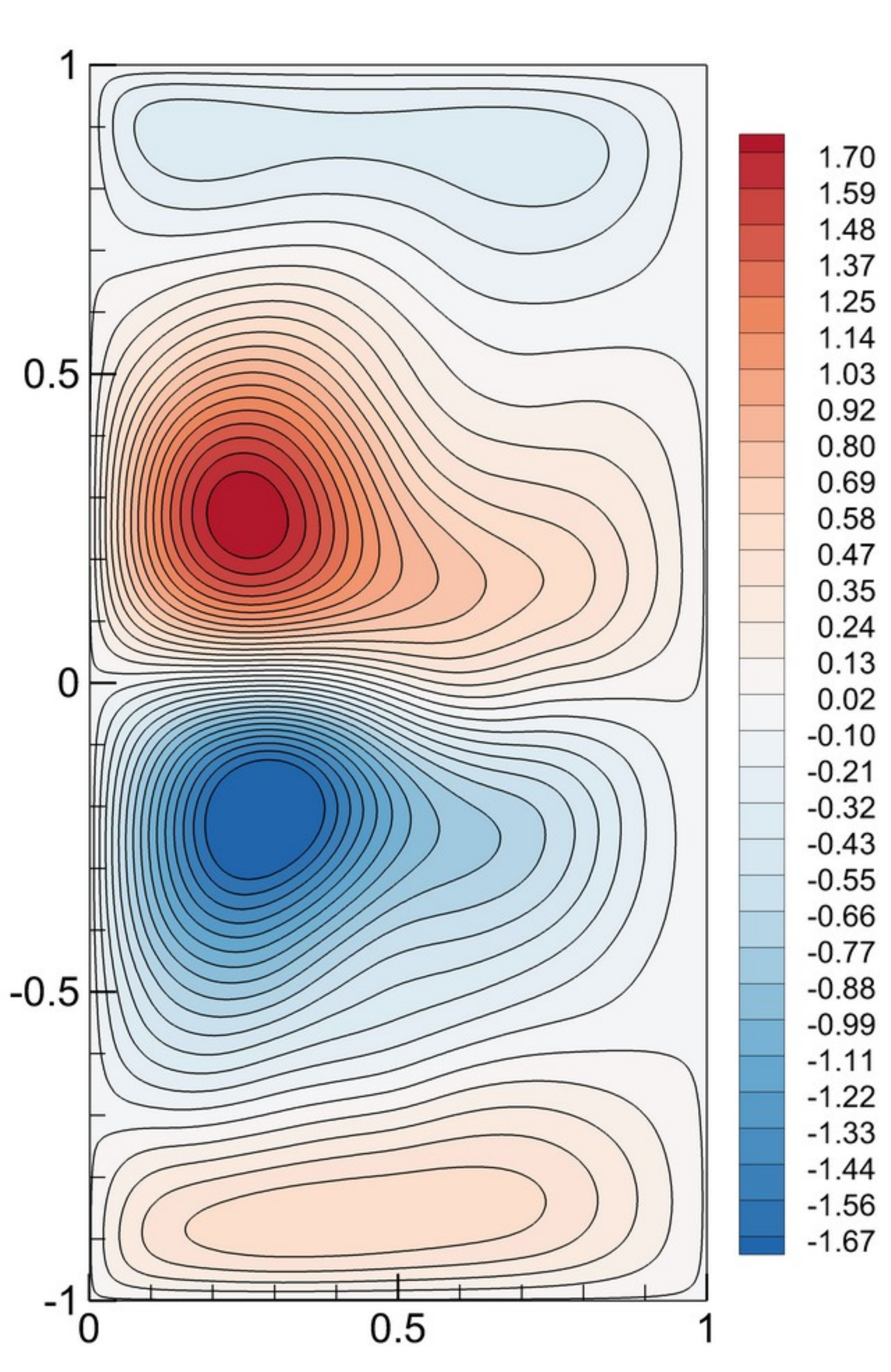}}
}
\caption{Comparison of the mean stream function contour plots for Experiment 2: (a) reference DNS computation at a resolution of $256 \times 512$ (with a computational cost of 324 hours of running CPU time); (b) standard Galerkin POD-ROM without any stabilization using $R=20$ modes (with a computational cost of 11 minutes of running CPU time); and (c) stabilized POD-ROM with $\nu_a=7.6$ using $R=20$ modes (with a computational cost of 11 minutes of running CPU time). Note that the standard Galerkin POD-ROM yields nonphysical result, whereas the DNS and the stabilized POD-ROM model results are qualitatively close. The contour interval layouts are identical only for (a) and (c).}
\label{fig:e2-com}
\end{figure}

\begin{table}
\caption{Characteristics of POD analysis for both experiments. The subscript $CF$ denotes the CPU time required to precompute POD coefficients given by Eqs.~(\ref{eq:roma1})-(\ref{eq:roma7}), and the subscript $ROM$ is the CPU time of the simulation with reduced-order model using the same $\Delta t$ of DNS. The CPU times for DNS are 326.03 h for Experiment 1 and 323.75 h for Experiment 2. The CPU time required to obtain POD basis functions by solving the eigensystem using the 700 snapshots is 2.31 h, including building the $C$ matrix. The superscript $G$ represents the Galerkin POD-ROM, and the superscript $S$ symbolizes the POD-ROM with the eddy viscosity stabilization scheme by using optimal $\nu_a$ values. The $L^2$-norms for the mean stream function are computed by using the data sets from the reference DNS.}
\label{tab:1}
\begin{tabular}{lllllll}
\hline\noalign{\smallskip}
R & $\sum\limits_{k=1}^{R}\lambda_{k}\Big/\sum\limits_{k=1}^{N}\lambda_{k}$ & CPU$_{CF}$ (s)  &
CPU$_{ROM}$ (s) & $\parallel \psi^{G}- \psi^{DNS}\parallel_{2}$ & $\parallel \psi^{S}- \psi^{DNS} \parallel_{2}$ & $\nu_a^{opt}$\\
\noalign{\smallskip}\hline\noalign{\smallskip}
\multicolumn{2}{l}{\textsl{Experiment 1}} \\
10 & 0.6159 & 8.75   & 85.43   & 262.0786 & 0.3013 & 6.6\\
20 & 0.7291 & 41.27  & 632.05  & 2.0049   & 0.4071 & 2.8\\
30 & 0.7855 & 112.23 & 2070.65 & 1.0036   & 0.4705 & 2.2\\
\multicolumn{2}{l}{\textsl{Experiment 2}} \\
10 & 0.4318 & 8.75   & 85.54   & 288.7832 & 0.4761 & 10.2\\
20 & 0.5825 & 41.29  & 632.50  & 87.0435  & 0.1463 & 7.6\\
30 & 0.6636 & 111.90 & 2070.01 & 41.0115  & 0.2045 & 5.8\\
\noalign{\smallskip}\hline
\end{tabular}
\end{table}

To illustrate the performance of the POD-ROM given by Eq.~(\ref{eq:rom1}), the four-gyre wind-driven circulation in a shallow ocean basin, a standard prototype of more realistic ocean dynamics, is considered. The model employs the BVE driven by a symmetric double-gyre wind forcing given by Eq.~(\ref{eq:forc}), which yields a four-gyre circulation in the time mean. This test problem has been used in numerous studies (e.g., \cite{cummins1992inertial,greatbatch2000four,nadiga2001dispersive,holm2003modeling,san2011approximate}). This problem represents an ideal test for the numerical assessment of the predictive performance of the POD-ROMs. Indeed, as showed in Greatbatch and Nadiga \cite{greatbatch2000four}, although a double gyre wind forcing is used, the long time average yields a four gyre pattern, which is challenging to capture on coarse spatial resolutions. As we will show in this study, this is also true for the POD-ROMs due to finite truncation in the reduced system. Thus, we will investigate numerically whether the new stabilized POD-ROM model can reproduce the four gyre time average using a small number of POD basis functions.

The mathematical model used in the four gyre problem is the BVE given by Eq.~(\ref{eq:nbve}). Following \cite{holm2003modeling,san2011approximate}, we utilize two different parameter sets, corresponding to two physical oceanic settings: Experiment 1 with a Rhines scale of $\delta_I/L = 0.06$ and a Munk scale of $\delta_M/L = 0.02$, which corresponds to a Reynolds number of $Re = 450$ and a Rossby number of $Ro = 0.0036$; and Experiment 2 with a Rhines scale of $\delta_I/L = 0.04$ and a Munk scale of $\delta_M/L = 0.02$, which corresponds to a Reynolds number of $Re = 200$ and a Rossby number of $Ro = 0.0016$. Since we set the Munk scale to $\delta_M/L = 0.02$ in our study, a grid resolution of $N_x>50$ in the $x$ direction ($L$ is the basin dimension in $x$ direction) represents the Munk layer resolving computation. Therefore, we use the Munk layer resolving computations as a reference solution which is denoted here as DNS. We emphasize that the term DNS in this study is not meant to indicate that a fully detailed solution is being computed on the molecular viscosity scale, but instead refers to resolving the simulation down to the Munk scale via the specified lateral eddy viscosity parameterization (e.g., see \cite{san2013approximate} for details). All numerical experiments conducted here are solved for a maximum dimensionless time of $T_{max} = 80$. This value corresponds to the dimensional times of 20.12 and 45.28 years for Experiment 1 and Experiment 2, respectively, which are long enough to capture statistically steady states.

To assess the POD-ROM, we employ the standard model reduction methodology. POD starts with data from an accurate numerical solution of underlying governing equation. The BVE equation is solved for the four-gyre problems by using the fourth-order Arakawa scheme for spatial derivatives and the third-order Runge-Kutta schemes for the time advancement process. We first run DNS computations on a fine mesh by using the resolution of $256\times512$ with a time step of $\Delta t = 2.5 \times 10^{-5}$. After a transient initial period, the flow exhibits a quasi-stationary regime for the present configurations. We store 700 snapshots in the time interval $[10, 80]$ at equidistant time intervals. We then build our POD-ROMs from these data sets.

In Fig.~\ref{fig:hist}, we plot the time evolution of the basin integrated total kinetic energy given by
\begin{equation}\label{eq:hist}
    E(t) = \frac{1}{2}\int\int \bigg(\Big(\frac{\partial \psi}{\partial x}\Big)^2 + \Big(\frac{\partial \psi}{\partial y}\Big)^2 \bigg) dxdy.
\end{equation}
We plot these time series for two parameter sets: Experiment 1 with $\delta_I/L = 0.06$ and $\delta_M/L = 0.02$,  and Experiment 2 with $\delta_I/L = 0.04$ and $\delta_M/L = 0.02$.  For both parameter sets, the time evolution of the above integral quantities follows the same pattern: after a short transient interval, they converge to the statistically steady state at a time of around $t=10$. Instantaneous contour plots at time $t=40$, $t=60$, and $t=80$ for the vorticity and stream function are shown in Fig.~\ref{fig:e1-ins} and Fig.~\ref{fig:e2-ins} for Experiment 1 and Experiment 2, respectively. The high variabilities of the flow dynamics in the statistically steady state are clearly seen in the plots showing no periodicity in this chaotic regime. Next, the POD data correlation matrix $C$ is constructed from the 700 snapshots between $t=10$ and $t=80$. Fig.~\ref{fig:eig} shows the eigenvalues of the correlation matrix $C$ for both experiments. The challenging nature of the problems can be seen from this plot. The eigenvalues here are slowly reduced in their amplitude by increasing the POD index.
%Therefore, it requires more and more POD modes to capture the essential dynamics of the flow field.
By looking at the slope of the eigenvalue distribution with respect to the POD index, it can be also seen that Experiment 2 is more challenging than Experiment 1 (i.e., the use of 30 modes in POD-ROM captures $78\%$ of the system's kinetic energy for Experiment 1 and $66\%$ of the energy for Experiment 2). Then, we construct the POD basis functions according to Eq.~(\ref{eq:basisw}) and Eq.~(\ref{eq:psi}). Some examples of corresponding POD modes for the stream function are shown in Fig.~\ref{fig:e1-bas} and Fig.~\ref{fig:e2-bas} for Experiment 1 and Experiment 2, respectively. It is clear that the smaller structures correspond to higher POD indices. The characteristics of POD analysis is also summarized in Table~\ref{tab:1}. The POD-ROM constructed using $R=10$ modes captures $62\%$ of the system's kinetic energy for Experiment 1, whereas it captures only $43\%$ for Experiment 2. As expected, the accuracy of the POD-ROMs increases with increasing $R$. In order to quantify the accuracies of the models, we compute the $L^2$-norms of the mean stream function errors with respect to the reference DNS data sets. In Table~\ref{tab:1}, $\parallel \psi^{G}- \psi^{DNS}\parallel_{2}$ shows the corresponding error norm for the Galerkin POD-ROM without using the eddy viscosity stabilization. It is shown that the error norm reduces from the amount of 262 for $R=10$ to the amount of 1 for $R=30$ in Experiment 1. A similar trend can be also seen in Experiment 2. However, due to its higher stiffness in Experiment 2, the error norm reduces from the amount of 289 for $R=10$ to the amount of 41 for $R=30$. We emphasize that the computational CPU time for the POD-ROM is considerable smaller than that for the actual DNS computations. %However, due to the quadratic nonlinearity for $R$ ordinary differential equations it is scaled as order of $R^3$. Thus, if we double the $R$ it requires about eight times more CPU time for the POD-ROM model. This effectively restricts the advantage of POD-ROM for smaller $R$ with a tradeoff in accuracy due to finite truncation error in the reduced-order model.

Next, we address the effects of stabilization for these complex convective flow settings. The sensitivity analysis with respect to the free stabilization parameter $\nu_a$ is systematically performed in what follows. Fig.~\ref{fig:e1-sens} and Fig.~\ref{fig:e2-sens} show the sensitivity anlayses for the Rempfer's mode dependent eddy viscosity stabilization given by Eq.~(\ref{eq:rem}). We compute the $L^2$-norms of the POD-ROMs' mean stream function error for $R=10$, $R=20$ and $R=30$. It can be noted that $\nu_a=0$ here corresponds to the standard Galerkin POD-ROM. It is clear that the effective stabilization considerably improves the accuracy of the POD-ROMs. The optimal values of the stabilization parameters are also shown in Table~\ref{tab:1}. The corresponding $L^2$-norms of the mean stream function errors are also listed in this table. The accuracy of the stabilized POD-ROM with $R=10$ is much higher than the accuracy of the standard POD-ROM with $R=30$. Considerably more accurate results are obtained by using the optimal values of eddy viscosity parameters. We also emphasize that the stabilization scheme presented in this paper is very efficient. It has a negligible computational overload due to linear viscosity kernel $k/R$ (the CPU time required for $R=10$ modes is 85.43 s for the stabilized POD-ROM, while it is 85.27 s for the standard POD-ROM).

We plot the time-averaged stream function contours in Fig.~\ref{fig:e1-sens} for Experiment 1. The stabilized POD-ROM model proposed in this study yields results that are significantly better than those corresponding to the standard Galerkin POD-ROM. Indeed, in the stream function plot in Fig.~\ref{fig:e1-sens} the stabilized POD-ROM model with $R=10$ clearly displays the correct four gyre pattern \cite{greatbatch2000four}. The standard Galerkin POD-ROM with the same number of POD modes incorrectly yields only two gyres with off values of stream functions, which is nonphysical. As shown in Fig.~\ref{fig:e2-sens}, similar observations can be made for the Experiment 2. It is clear that the proposed POD-ROM model yields accurate results that are close to the DNS results and has a considerably reduced computational cost. As reported in Table~\ref{tab:1}, the CPU times for DNS are 326.03 h for Experiment 1 and 323.75 h for Experiment 2. The CPU time required to obtain the POD basis functions by building the $C$ matrix and solving the eigensystem using the 700 snapshots takes around 2.31 h. After all the POD-ROM coefficients are precomputed, both the standard and the stabilized POD-ROMs run efficiently on the order of seconds.

\section{Summary and Conclusions}
\vspace{-2pt}
\label{sec:conc}
A stabilized POD-ROM for the BVE modeling the large scale flows in quasigeostropic systems was presented. The POD-ROM was tested in the numerical simulation of the wind-driven circulation in a shallow ocean basin, a standard prototype of more realistic ocean dynamics, where a symmetric double-gyre wind forcing yields four-gyre circulations in the time mean.  To reduce the error associated with the numerical discretization, we used a fourth-order Arakawa scheme for the spatial discretization and a third-order Runge-Kutta scheme for the temporal integration. %Utilizing the method of snapshots along with the Galerkin projection, the various reduced-order models were derived in reduced subspaces.
Two numerical examples were used to assess the performance of the POD-ROMs. We constructed different ROMs with different numbers of POD basis functions and stabilization parameters. Results obtained from these POD-ROMs were compared with those calculated by DNS.

We showed that the large scale quasigeostropic dynamics are well captured by the proposed POD-ROM even with a few modes. The stabilized POD-ROM yielded numerical results that were in close agreement with those of the DNS. In particular, the four gyre structure of the time-averaged stream function contour plots was recovered by the proposed POD-ROM. Using the same number of modes as the stabilized POD-ROM, the standard Galerkin POD-ROM %without any stabilization model on the same number of POD modes as that employed by the stabilized model
produced inaccurate, unphysical results. We also performed a numerical investigation of the sensitivity with respect to the free stabilization parameters used and we found that the stabilized POD-ROM is robust. This first step in the numerical assessment of the proposed POD-ROM shows that it
%model shows that the POD-ROM presents very reliable low-dimensional model and
could represent a viable model reduction tool in numerical weather prediction and climate modeling.
%systems as well as data assimilation purposes.

%\begin{acknowledgements}
%\end{acknowledgements}

\bibliographystyle{spmpsci}      % mathematics and physical sciences
\bibliography{ref}   % name your BibTeX data base

%\begin{thebibliography}{}
% etc
%\end{thebibliography}

\end{document}